\documentclass[]{aa}  
\usepackage{color }
\usepackage{natbib}
\usepackage{amsmath}
\usepackage{amssymb}
\usepackage{graphicx}
\usepackage{txfonts}
\usepackage{soul}
\newcommand{\bl}[1]{\mbox{\boldmath$ #1 $}}
\begin{document}

\title{The effect of luminosity outbursts on the protoplanetary disk dynamics}
\titlerunning{Luminosity bursts affect disk dynamics}

\author{Eduard I. Vorobyov\inst{1,2,3}, Vardan G. Elbakyan\inst{2}, Michihiro Takami\inst{4}, and Hauyu Baobab Liu\inst{4}}
\authorrunning{Vorobyov et al.}
\institute{ 
University of Vienna, Department of Astrophysics, Vienna, 1180, Austria \\
\email{eduard.vorobiev@univie.ac.at} 
\and
Research Institute of Physics, Southern Federal University, Rostov-on-Don, 344090 Russia
\and
Ural Federal University, 51 Lenin Str., 620051 Ekaterinburg, Russia
\and
Institute of Astronomy and Astrophysics, Academia Sinica, 11F of Astronomy-Mathematics Building, AS/NTU No.1, Sec. 4, Roosevelt
Rd, Taipei 10617, Taiwan, R.O.C.
}

  \abstract
  {}
   {Response of a protoplanetary disk to luminosity bursts of various duration is studied with the purpose to determine the effect of the bursts on the strength and sustainability of gravitational instability in the disk. A special emphasis is paid to the spatial distribution of gas and grown dust ({from 1 mm to a few cm}) during and after the burst.}
   {Numerical hydrodynamics simulations using the FEOSAD code were employed to study the dynamics of gas and dust in the thin-disk limit. Dust-to-gas friction including back reaction and dust growth were also considered. Bursts of various duration (from 100 to 500~yr) were initiated  in accordance with a thermally ignited magnetorotational instability. Luminosity curves for constant- and declining-magnitude bursts were adopted to represent two typical limiting cases for FU-Orionis-type eruptions.}
{The short-term effect of the burst is to reduce the strength of gravitational instability by heating  and expanding the disk. The longest bursts  with duration comparable to the revolution period of the spiral can completely dissolve the original two-armed spiral pattern in the gas disk by the end of the burst, while the shortest bursts only weaken the spiral pattern. The reaction of grown dust to the burst is somewhat different. The spiral-like initial distribution with deep cavities in the inter-armed regions transforms into a ring-like distribution with deep gaps. This transformation is most expressed for the longest-duration bursts. 
The long-term effect of the burst depends on the initial disk conditions at the onset of the burst. In some cases, vigorous disk fragmentation sets in several thousand years after the burst, which was absent in the model without the burst. Several clumps with masses in the giant-planet mass range form in the outer disk regions. After the disk fragmentation phase, the spatial distribution of grown dust is characterized by multiple sharp rings located from tens to hundreds of astronomical units.  The arrangement and sharpness of the rings depends on the strength of dust turbulent diffusion. The wide-orbit rings are likely formed as the result of dust-rich clump dispersal in the preceding gravitationally unstable phase.}
 {Luminosity bursts similar in magnitude to FU-Orionis-type eruptions can have a profound effect on the dynamics of gas and dust in protoplanetary disks  if the burst duration is comparable to or longer than the dynamical timescales. In this context, the spatial morphology of the gas-dust disk of V883~Ori, a FUor-like object that is thought to be in the outburst phase for more than a century with the unknown onset date, may be used as test case for the burst models considered here. A potential relation of the obtained ring structures to a variety of gaps and rings observed in T~Tauri disks remains to be established.
   }

   \keywords{Protoplanetary disks -- Stars: protostars -- instabilities }

   \maketitle

\section{Introduction}
Protostellar accretion is a fundamental process in the theory of star formation. Accretion sets the terminal mass of protostars, while radiative energy released by accretion provides a notable contribution to the total luminosity in the early stages of evolution \citep{2016Elbakyan}, thus influencing appreciably the thermal balance of circumstellar disks. The manner in which protostellar accretion evolves with time -- constant, steadily declining, or variable -- is still under debate. 

Both observational and theoretical evidence suggest that protostellar accretion can feature violent bursts known as FU-Orionis-type eruptions (FUors). The accretion rate during these events can rise by several orders of magnitude up to $10^{-4}~M_\odot$~yr$^{-1}$ and the corresponding accretion luminosity can amount to several hundreds of solar luminosity \citep{2014AudardAbraham}. Such a dramatic increase in luminosity is known to have a profound effect on the disk chemistry \citep{2007Lee}. The accompanying rise in the disk temperature leads to sublimation of icy mantles from dust grains, this shifting the ice lines \citep{2016Cieza}. Luminosity bursts can also trigger the chemical gas-phase reactions that are typically dormant in the quiescent phases, leading to long-lasting changes in the disk chemical composition \citep{2019Wiebe}.  

The profound effect of luminosity bursts on the disk chemical evolution can be accompanied by notable changes in the dynamical evolution of the entire disk. \citet{2011Stamatellos} found that luminosity bursts can suppress disk fragmentation, but the process resumes in between the bursts during longer periods of quiescent accretion. A similar phenomenon is observed when a secondary companion (and not the primary star) experiences a luminosity burst \citep{2017MercerStamatellos}. Moreover, preferential freeze-out of volatile species on dust grains after the burst can affect dust growth in protoplanetary disks \citep{2017Hubbard}.

In this work, we study numerically the effect of luminosity bursts on the dynamical evolution of gas-dust disks. Unlike \citet{2011Stamatellos}, we pay specific attention to the duration of luminosity bursts, a characteristic that is poorly constrained from observations.  All known FUors, perhaps, with the exception of V346~Nor \citep{2020Kospal}, are still in the active burst phase. This makes the longest confirmed timescale for the burst approaching 100~yr for FU~Orionis and this system shows no sign of abating.
Furthermore, analysis of photographic plates suggests that V883~Ori was already in outburst in 1890 \citep{1890Pickering}, making this source the longest known FUor. Numerical models also predict a wide range of burst durations from a few tens to several hundred years \citep[see fig. 7 in][]{2014AudardAbraham}.

The case of V883~Ori is particularly interesting in this context because of its rather high disk mass of $\ga 0.3~M_\odot$ \citep{2018Cieza}, for which theoretical arguments suggest gravitational instability and even fragmentation \citep{2013Vorobyov}. Moreover, V883 Ori does not have an obvious companion or a close-orbit planet, suggesting that the burst must have been caused  by other mechanisms that imply the presence of gravitational instability in the disk \citep{2014BaeHartmann,VorobyovBasu2015}.
If V883~Ori is more than 100~yr in the outburst, the original spatial morphology of its disk may have already been altered.

Here, we explore this possibility in detail by setting luminosity bursts of different duration in systems featuring gravitationally unstable disks. We then follow the evolution of circumstellar disks for several tens of kyr and compare the burst cases with the evolution of the disk without the burst. For the first time, we explore the effect of the burst not only on the gaseous, but also on the dusty component of the disk. For this purpose, we employ the FEOSAD numerical hydrodynamics code in the thin-disk limit presented in \citet{2018VorobyovAkimkin}. The paper is organized as follows. In Sect.~\ref{SCS} we briefly describe the numerical model. In Sect.~\ref{Global} we present the global disk evolution without the burst. 
In Sect.~\ref{short_burst} the disk response to the burst is considered. In Sect.~\ref{long_burst} we consider the long-term effect (several tens of kyr) of the bursts of various duration on the development of gravitational instability and fragmentation in the disk.  The main results are summarized in Sect.~\ref{Summary}.


\section{Model description}
\label{SCS}
We use the FEOSAD (Formation and Evolution Of Stars and Disks) numerical hydrodynamics code to model the evolution of a protoplanetary gas-dust disk in the thin-disk limit \citep{2018VorobyovAkimkin}. Numerical simulations start from the gravitational collapse of a pre-stellar core with a mass of $M_{\rm core}=0.59~M_\odot$ and ratio of rotational-to-gravitational energy of $\beta=0.24\%$. These parameters were chosen to produce a sufficiently massive disk prone to gravitational instability, but stable to fragmentation. 
The initial dust to gas mass ratio was set equal to 0.01. The central star is modelled by a point gravitating source, which grows in mass through accretion from the circumstellar disk. The physical model includes disk self-gravity, stellar and background irradiation, disk radiative cooling, viscous heating, dust growth, and dust-to-gas friction including back reaction of dust on gas. 


\subsection{Gas component}
The hydrodynamic equations of mass, momentum, and internal energy for the gas component read
\begin{equation}
\label{cont}
\frac{{\partial \Sigma_{\rm g} }}{{\partial t}}   + \nabla_p  \cdot 
\left( \Sigma_{\rm g} \bl{v}_p \right) =0,  
\end{equation}
\begin{eqnarray}
\label{mom}
\frac{\partial}{\partial t} \left( \Sigma_{\rm g} \bl{v}_p \right) +  [\nabla \cdot \left( \Sigma_{\rm
g} \bl{v}_p \otimes \bl{v}_p \right)]_p & =&   - \nabla_p {\cal P}  + \Sigma_{\rm g} \, \bl{g}_p + \nonumber
\\ 
&+& (\nabla \cdot \mathbf{\Pi})_p  - \Sigma_{\rm d,gr} \bl{f}_p,
\end{eqnarray}
\begin{equation}
\frac{\partial e}{\partial t} +\nabla_p \cdot \left( e \bl{v}_p \right) = -{\cal P} 
(\nabla_p \cdot \bl{v}_{p}) -\Lambda +\Gamma + 
\left(\nabla \bl{v}\right)_{pp^\prime}:\Pi_{pp^\prime}, 
\label{energ}
\end{equation}
where the subscripts $p$ and $p^\prime$ refer to the planar components
$(r,\phi)$  in polar coordinates, $\Sigma_{\rm g}$ is the gas mass
surface density,  $e$ is the internal energy per surface area,  ${\cal P}$
is the vertically integrated gas pressure calculated via the ideal  equation of state as ${\cal P}=(\gamma-1) e$ with $\gamma=7/5$, $f_p$ is the friction force between gas and dust, $\bl{v}_{p}=v_r
\hat{\bl r}+ v_\phi \hat{\bl \phi}$  is the gas velocity in the disk plane, and is $\nabla_p=\hat{\bl r} \partial / \partial r + \hat{\bl
\phi} r^{-1} \partial / \partial \phi $ the gradient along the planar coordinates of the disk.  

The gravitational acceleration in the disk
plane,  $\bl{g}_{p}=g_r \hat{\bl r} +g_\phi \hat{\bl \phi}$, takes into account disk self-gravity found by solving for the Poisson integral \citep{1987Binney} and the
gravity of the central protostar when formed. Turbulent viscosity is
taken into account via the viscous stress tensor  $\mathbf{\Pi}$, the expression for which can be found in \citet{2010VorobyovBasu}. We parameterized the
magnitude of kinematic viscosity $\nu=\alpha c_{\rm s} H$  using the $\alpha$-prescription of \citet{1973ShakuraSunyaev}, where $c_{\rm s}$ is the sound speed  and $H$ is the vertical scale height of the gas disk calculated using an assumption of local hydrostatic equilibrium. The value of $\alpha=0.01$ is a constant of time and space.

The expressions for radiative cooling and irradiation heating read \citep{2016DongVorobyov},
\begin{equation}
\Lambda=\frac{8\tau_{\rm P} \sigma T_{\rm mp}^4}  {1+2\tau_{\rm P} + 
{3 \over 2}\tau_{\rm R}\tau_{\rm P}},
\end{equation}
\begin{equation}
\Gamma=\frac{8\tau_{\rm P} \sigma T_{\rm irr}^4 }{1+2\tau_{\rm P} + {3 \over 2}\tau_{\rm R}\tau_{\rm
P}},
\end{equation}
where $\tau_{\rm P}=\kappa_{\rm P} \Sigma_{\rm d,tot}$ and $\tau_{\rm R}=\kappa_{\rm R} \Sigma_{\rm d,tot}$ are the Planck and Rosseland optical depths, $\sigma$ is the Boltzmann constant, $T_{\rm mp}$ is the midplane hydrodynamic temperature, and $T_{\rm irr}$ is the  temperature of stellar and background radiation.The Rosseland and Planck opacities ($\kappa_{\rm P}$ and $\kappa_{\rm R}$) were taken from \citet{2003SemenovHenning}. The total dust surface density $\Sigma_{\rm d,tot}$ is found from solving the hydrodynamic equations for the dust component as described below.  We note that the cooling and heating rates in \citet{2016DongVorobyov}
were written for one side of the disk and need to be multiplied by a factor of 2. 

The irradiation temperature is composed of the inputs from background and stellar radiation and is defined as 
\begin{equation}
T_{\rm irr}^4=T_{\rm bg}^4+\frac{F_{\rm irr}(r)}{\sigma},
\label{fluxCS}
\end{equation}
where $T_{\rm bg}$ is the temperature of background irradiation set equal to the initial temperature of prestellar core (20~K) and $F_{\rm
irr}(r)$ is the radiation flux   absorbed by the disk surface at radial distance  $r$ from the
central star. The latter quantity is calculated as 
\begin{equation}
F_{\rm irr}(r)= \frac{L_\ast}{4\pi r^2} \cos{\gamma_{\rm irr}},
\label{fluxF}
\end{equation}
where $\gamma_{\rm{irr}}$ is the incidence angle of radiation arriving at
the disk surface (with respect to the normal) at radial distance $r$ calculated as in \citet{2010VorobyovBasu}. The stellar luminosity $L_{\ast}$ is the sum of the
accretion luminosity  $L_{\rm {\ast,accr}}=0.5 G M_{\ast} \dot{M}
/R_{\ast}$ arising from the gravitational energy of accreted gas and the
photospheric luminosity $L_{\rm{\ast,ph}}$ due to gravitational
compression and deuterium burning in the stellar interior. The stellar
mass $M_{\ast}$ and accretion rate onto the star $\dot{M}$ are determined
using the amount of gas accreted by the star. 
The properties of
the forming protostar ($L_{\rm{\ast,ph}}$ and radius $R_{\ast}$) are
calculated using the stellar evolution tracks obtained with the STELLAR code of \citet{2008YorkeBodenheimer}. We note that in our thin-disk model the disk is assumed to be vertically (but not radially or azimuthally) isothermal. 

\subsection{Dust component}
\label{DustComp}
In our model, dust consists of two components: small dust ($a_{\rm min}<a<a_\ast$)  and grown dust ($a_\ast \le a<a_{\rm max}$), where $a_{\rm min}=5\times 10^{-3}$~$\mu$m, $a_\ast=1.0$~$\mu$m,  and $a_{\rm max}$ is a dynamically varying maximum radius of dust grains, which depends on the efficiency of radial dust drift and on the rate of dust growth. All dust grains are considered to be spheres with material density $\rho_{{\rm s}}=2.24\,{\rm g~cm}^{-3}$. Small dust constitutes the initial reservoir
for dust mass and is gradually converted in grown dust as the disk forms and evolves.   Small dust is assumed to be dynamically coupled to gas, meaning that we only solve the continuity equation for small dust grains, while the dynamics of grown dust is controlled by friction with the gas component and by the total gravitational potential of the star, as well as the gaseous and dusty components.
The continuity and momentum equations for small and grown dust are
\begin{equation}
\label{contDsmall}
\frac{{\partial \Sigma_{\rm d,sm} }}{{\partial t}}  + \nabla_p  \cdot 
\left( \Sigma_{\rm d,sm} \bl{v}_p \right) = - S(a_{\rm max}),  
\end{equation}
\begin{equation}
\label{contDlarge}
\frac{{\partial \Sigma_{\rm d,gr} }}{{\partial t}}  + \nabla_p  \cdot 
\left( \Sigma_{\rm d,gr} \bl{u}_p \right) = S(a_{\rm max}),  
\end{equation}
\begin{eqnarray}
\label{momDlarge}
\frac{\partial}{\partial t} \left( \Sigma_{\rm d,gr} \bl{u}_p \right) +  [\nabla \cdot \left( \Sigma_{\rm
d,gr} \bl{u}_p \otimes \bl{u}_p \right)]_p  &=&   \Sigma_{\rm d,gr} \, \bl{g}_p + \nonumber \\
 + \Sigma_{\rm d,gr} \bl{f}_p + S(a_{\rm max}) \bl{v}_p,
\end{eqnarray}
where $\Sigma_{\rm d,sm}$ and $\Sigma_{\rm d,gr}$ are the surface
densities of small and grown dust and $\bl{u}_p$ describes the planar components of the grown dust velocity.

The rate of small-to-grown dust conversion $S(a_{\rm max})$ is derived based on the assumption that each of the two dust populations (small and grown) has the size distribution $N(a)$ described by a simple power-law function $N(a)= C a^{-p}$ with a fixed exponent $p=3.5$ and a normalisation constant $C$. After noting than that total dust mass stays constant during dust growth, the change in the surface density of small dust due to conversion into grown dust $\Delta \Sigma_{\rm d,sm}=\Sigma_{\rm d,sm}^{n+1}-\Sigma_{\rm d,sm}^{n}$ can be expressed as
\begin{equation}
    \Delta\Sigma_{\mathrm{d,sm}} =  \Sigma_{\mathrm{tot}}^n 
    \frac
    { I_1
    \left( 
    C_{\rm sm}^{n+1} C_{\rm gr}^n \, I_2 - 
    C_{\rm sm}^n C_{\rm gr}^{n+1} I_3
    \right)
    }
    {
    \left( 
    C_{\rm sm}^{n+1} I_1+  C_{\rm gr}^{n+1} I_3
    \right)  
    \left( 
    C_{\rm sm}^{n} I_1 + C_{\rm gr}^{n} I_2
    \right)
    \label{growth1}
    },
\end{equation}
where $C_{\rm sm}$ and $C_{\rm gr}$ are the normalisation constants for small and grown dust size distributions at the current ($n$) and next ($n+1$) time steps,   and the integrals $I_1$, $I_2$, and $I_3$ are defined as
\begin{equation}
   I_1= \int_{a_{\rm min}}^{a_*} a^{3-\mathrm{p}}da, \,\,\,\,
    I_2=\int_{a_*}^{a_{\mathrm{max}}^{n}} a^{3-\mathrm{p}}da,
    \,\,\,\,
    I_3=\int_{a_*}^{a_{\mathrm{max}}^{n+1}} a^{3-\mathrm{p}}da.
\end{equation}
We further assume that $C^n_{\rm sm}=C^{n}_{\rm gr}$ and $C^{n+1}_{\rm sm}=C^{n+1}_{\rm gr}$, allowing us to cancel out the normalization constants in Equation~(\ref{growth1}). This assumption implies that the dust size distribution is continuous across $a_\ast$. The cases with a discontinuous dust size distributions will be considered in follow-up studies. Finally, the dust growth rate can be written 
\begin{equation}
 S(a_{\rm max}) = - {\Delta\Sigma_{\mathrm{d,sm}} \over \Delta t},   
\end{equation}
where $\Delta t$ is the hydrodynamic time step.

To complete the calculation of $S(a_{\rm max})$, the maximum radius of grown dust $a_{\rm max}$ in a given computational cell must be computed at each time step. 
The evolution of $a_{\rm max}$ is described as
\begin{equation}
{\partial a_{\rm max} \over \partial t} + (u_{\rm p} \cdot \nabla_p ) a_{\rm max} = \cal{D},
\label{dustA}
\end{equation}
where the growth rate $\cal{D}$ accounts for the change in $a_{\rm max}$ due to coagulation and the second term on the left-hand side account for the change of $a_{\rm max}$ due to dust flow through the cell.  We write the source term $\cal{D}$ as
\begin{equation}
\cal{D}=\frac{\rho_{\rm d} {\it v}_{\rm rel}}{\rho_{\rm s}},
\label{GrowthRateD}
\end{equation}
where $\rho_{\rm d}$ is the total dust volume density and $v_{\rm rel}$ is the
dust-to-dust collision  velocity. 
The adopted approach is similar to 
the monodisperse model of \citet{1997Stepinski} and is described in more detail in \citet{2018VorobyovAkimkin}.
To summarise, our dust growth model allows us to determine the maximum size of dust grains, assuming that the slope $p$ of the dust size distribution is continuous and stays locally and globally constant.
The dust growth is limited by the fragmentation barrier defined as \citep{2012Birnstiel}
\begin{equation}
 a_{\rm frag}=\frac{2\Sigma_{\rm g}v_{\rm frag}^2}{3\pi\rho_{\rm s}\alpha c_{\rm s}^2}
 \label{afrag}
\end{equation}
and the fragmentation velocity is set to $v_{\rm farg}=30$~m~s$^{-1}$, which corresponds to sticky icy grains \citep{2013Wada, 2019Charnoz}.

As dust grows, the span in the dust sizes  and in the corresponding Stokes numbers covered by the grown component may become substantial. However, we only track the dynamics of dust particles with the maximum size by calculating the stopping time for $a_{\rm max}$.  For the chosen power law of $p=3.5$, most of the dust mass is located at the upper end of the dust size distribution, meaning that the dust particles of maximum size are representative for the entire dust mass reservoir. A more rigorous approach  to studying dust dynamics requires the use of multi-bins with narrower ranges of dust sizes and this model is currently under development.

\subsection{Boundary conditions and solution method}
The simulations were performed on the polar grid ($r,\phi$) with $512\times 512$ grid cells. The radial grids are logarithmically spaced, while the azimuthal grids have equal spacing. The central disk region of 1.0~au in radius is carved out and replaced with the sink cell to avoid too small time steps imposed by the Courant conditions. 
Care should be taken when choosing the type of flow through the inner boundary.  If
the inner boundary allows for matter to flow only in
one direction, i.e., from the disk to the sink cell, then
any wave-like motions near the inner boundary, such as
those triggered by spiral density waves in the disk, would
result in a disproportionate flow through the sink--disk
interface. As a result, an artificial depression in the gas
density near the inner boundary develops in the course
of time because of the lack of compensating back flow
from the sink to the disk.

To reduce the possible negative effect of the disproportionate flow, the FEOSAD code features the special inflow-outflow inner boundary condition at the sink cell--disk interface as described in \citet{2019VorobyovSkliarevskii}. To proceed with numerical simulations, the values of surface densities, velocities, and gas internal energy at the computational boundaries need to be defined. The simplest practice is to assume zero gradients across the inner boundary. We, however, found that this leads to an artificial drop in the gas surface density near the disk--sink interface as noted above. We developed the central smart cell (CSC) model, which allowed us to greatly reduce the artificial density drop (and often even eliminate it completely) and also account for MRI-triggered bursts in the innermost disk regions. 

In the CSC model, the surface density at the boundary is computed from a system of mass balance equations. We ensure that the model conserves the total mass budget in the system, but the momentum and energy in the CSC are not computed.  Instead, the radial velocity and internal energy at the inner boundary
are determined from the zero gradient condition. The azimuthal velocity is extrapolated from the active disk to the sink cell assuming a Keplerian rotation. A more rigorous approach involves solving for one-dimensional system of hydrodynamic equations in the sink cell with a  complicated interface between the sink and the active disk, like in \citet{2007Crida}. This approach my be undertaken in future updates of the FEOSAD code.

The balance of mass in the CSC is computed using the following system of ordinary differential equations:
\begin{equation}
{dM_{\rm csc} \over{dt} }= \dot{M}_{\rm disk}-\dot{M}_{*, \rm csc}-\dot{M}_{*, \rm bst}-\dot{M}_{\rm jet}
\label{csc1}
,\end{equation}
\begin{equation}
\label{csc2}
{dM_{*} \over{dt} } = \dot{M}_{*, \rm csc}+\dot{M}_{*,\rm  bst},
\end{equation}
where $M_{\rm csc}$ is the mass of gas and dust in the CSC and $M_\ast$ is the mass of the star.
Here, $\dot{M}_{\rm disk}$ is the mass accretion rate through the CSC--disk interface calculated   as the mass flux passing through the inner disk boundary and  $\dot{M}_{\rm jet}$ is the mass ejection rate by jets and outflows, taken to be 10\% that of $\dot{M}_{\rm disk}$. The mass accretion rate from the CSC on the star is split into two parts: $\dot{M}_{*, \rm csc}$ reflecting the regular mode of mass accretion from the CSC to the star  and $\dot{M}_{*, \rm bst}$  denoting the burst mode of accretion caused by the thermally ignited MRI in the CSC. The latter occurs if the gas temperature at the CSC--disk interface exceeds a threshold value $T_{\rm crit}$ for thermal ionization of alkaline metals to set in. Initially, the value of $T_{\rm crit}$ is set equal to 1600~K. We note that this is a rather high value \citep{2015Desch} and with this choice our model features only occasional MRI bursts in the very early disk evolution ($t<100$~kyr). We did it intentionally to produce a clean run that minimizes the likelihood of unexpected bursts. In the subsequent evolution, when we want to make a controlled experiment  to study the disk response to the burst, we simply lower $T_{\rm crit}$ to a more realistic value of 1300~K, thus initiating a burst at a time instance of our choice (see Sect.~\ref{short_burst}).


We assume that regular (non-burst) mass accretion from the CSC on the star is a fraction $\xi$ of mass accretion from the disk to the CSC
\begin{equation}
\dot{M}_{*, \rm csc}=
\begin{cases}
\xi \dot{M}_{\rm disk} &\text{for $\dot{M}_{\rm disk} > $ 0}, 
\label{Sink:1}
\\ 
0 &\text{for $\dot{M}_{\rm disk} \leqslant $ 0}.
\end{cases}
\end{equation}
The limit of small $\xi  \approx 0$ corresponds to slow mass transport through the CSC, resulting in gradual mass accumulation in the sink. Slow mass transport may be caused by the development of a dead zone in the CSC. The opposite limit of large $\xi \approx 1.0$ assumes fast mass transport, so that the matter that crosses the CSC--disk interface does not accumulate in the CSC, but is quickly delivered on the star. Physically, this means that mass transport mechanisms of similar efficiency operate in the disk and in the CSC. In this work, $\xi$ is set equal to 0.05. Slow mass transport leads to mass accumulation and gas heating in the CSC and its immediate vicinity, thus allowing MRI bursts. We emphasize that the CSC is not exclusively introduced to model the bursts but is an integral part of the FEOSAD code, which allows us to better simulate the sink--disk interface. 

\begin{figure}
\begin{centering}
\includegraphics[width=1\columnwidth]{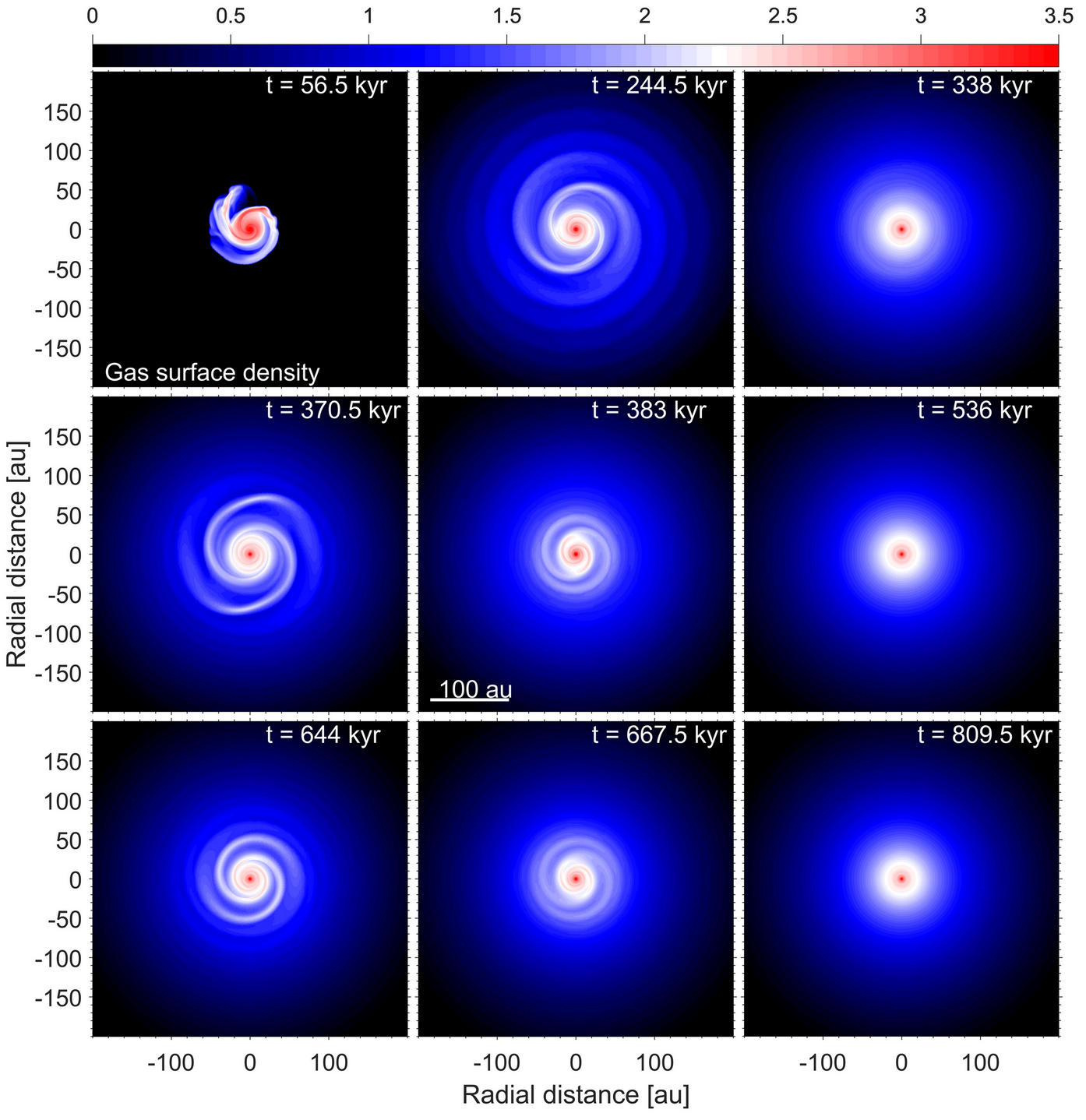}
\par\end{centering}
\caption{\label{fig:1} Temporal evolution of the gas component of the disk. Shown are nine time instances from the onset of numerical simulations. The disk forms at $t=34$~kyr and the embedded phase ends at $t=91$~kyr. The scale bar shows the gas surface density in g~cm$^{-2}$ (log units).}
\end{figure}

 Equations (\ref{cont})--(\ref{energ}), (\ref{contDsmall})--(\ref{momDlarge}),  and (\ref{GrowthRateD}) are solved using
the operator-split solution procedure as described in the
ZEUS code \citep{1992StoneNorman}. The solution is
split in the transport and source steps. In the transport
step, the update of hydrodynamic quantities due to advection
is done using the third-order piecewise parabolic interpolation scheme of \citet{1984Colella}. This step
also considers the change of maximum dust radius due to
advection. The FARGO algorithm is also used to ease the strict limitations on the
Courant condition, which occur in numerical simulations
of Keplerian disks using the curvilinear coordinate systems
converging towards the origin \citep{2000Masset}. To account for the friction
terms, we apply the analytic solution as described in \citet{2018Stoyanovskaya}.

\section{Global disk evolution}
\label{Global}
We begin with showing in Figure~\ref{fig:1} the global disk evolution over a time period of 810~kyr. The disk forms at $t=34$~kyr since the onset of cloud core collapse. Twenty thousand years later the disk already exhibits a well-defined spiral pattern, indicating the development of gravitational instability.  Interestingly, the disk experiences several episodes of gravitational instability. The first episode occurs in the embedded stage of disk evolution when the disk is subject to intense mass-loading from the infalling envelope. The spiral structure is initially rather irregular and lopsided, but gradually evolves into a regular two-armed spiral pattern and finally dissolves at $t\approx330$~kyr, although a week one-armed spiral can still be discerned. The second episode of disk destabilization takes place around $t=370$~kyr, already in the early T~Tauri stage when most of the envelope material has accreted on the disk plus star system. The spiral structure is dominated by a two-armed pattern. The disk stabilizes again after $t=500$~kyr. The final episode of disk destabilization (at least during the computed period of time) occurs around $t=640$~kyr and continues for several tens of thousands of years. The disk never experiences fragmentation. The disk and stellar masses at the end of our numerical simulations are $0.14~M_\odot$ (inside 100~au) and $0.32~M_\odot$, respectively.  Ten per cent of the initial core mass ($0.59~M_\odot$) was ejected with the jets (see Eq.~\ref{csc1}) and the rest resides in a diffuse outer disk and envelope outside of 100~au.

We can visualize the cycles of disk destabilization in terms of the global Fourier amplitudes, which measure the strength of gravitational instability and can be defined as 
\begin{equation}
C_{\rm m} (t) = {1 \over M_{\rm d}} \left| \int_0^{2 \pi} 
\int_{r_{\rm csc}}^{R_{\rm d}} 
\Sigma_{\rm g}(r,\phi,t) \, e^{im\phi} r \, dr\,  d\phi \right|,
\label{fourier}
\end{equation}
where $R_{\rm in}=10$~au and $R_{\rm out}=80$~au are the inner and outer disk radii encompassing the localization of the spiral pattern, $M_{\rm d}$ is the mass of the disk within this region, $m$ is the ordinal number of the spiral mode, and $\Sigma_{\rm g}$ is the gas surface density.  When the disk surface density is axisymmetric, the amplitudes of all modes are equal to zero. When, say, $C_{\rm m}(t)=0.01$, the perturbation amplitude of  spiral density waves in the disk is 1\% that of  the underlying axisymmetric density distribution. 

\begin{figure}
\begin{centering}
\includegraphics[width=1\columnwidth]{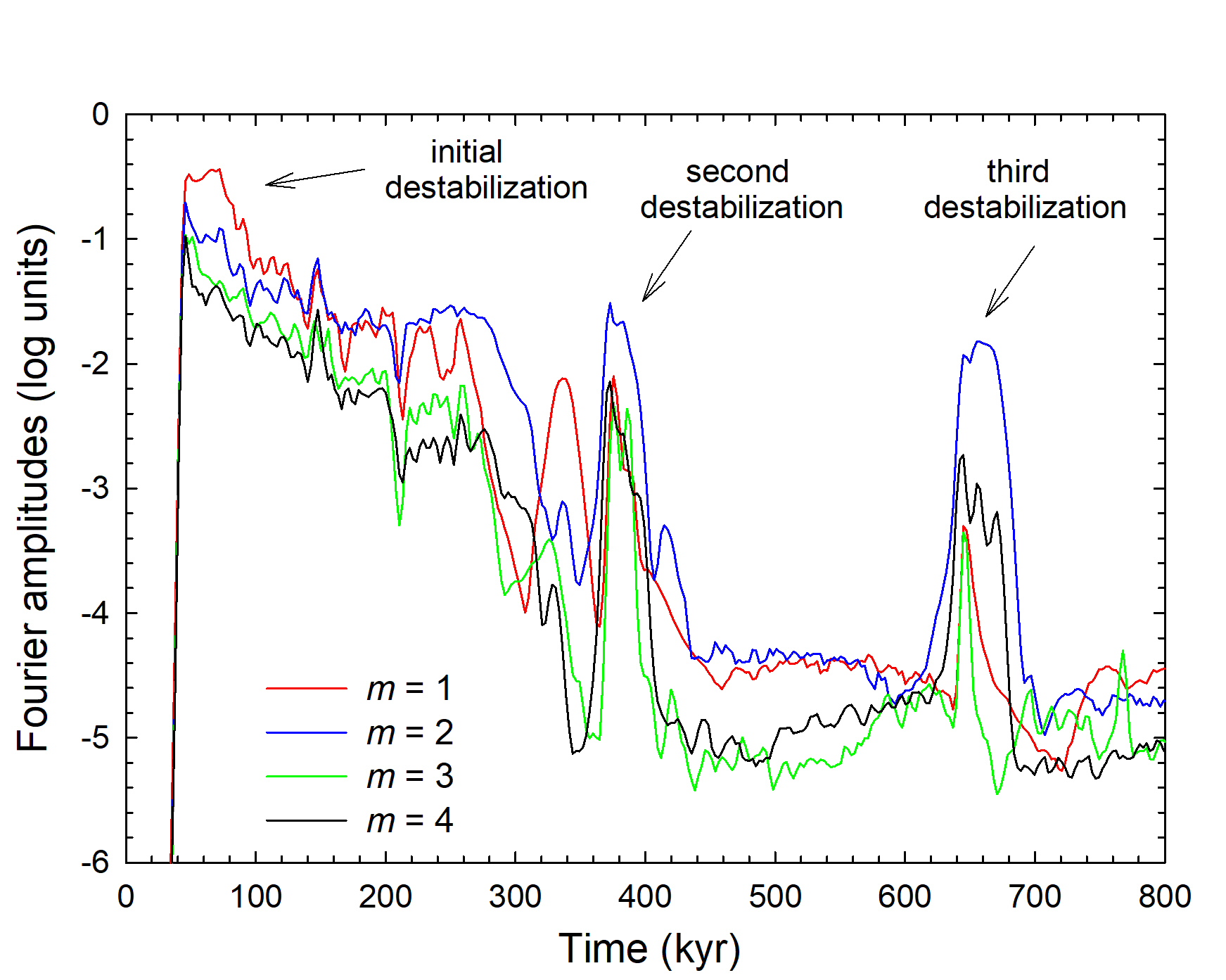}
\par\end{centering}
\caption{\label{fig:2} Global Fourier amplitudes of the disk (in log units) as a function of time. Shown are amplitudes for the initial four modes $m$=1,2,3, and 4.}
\end{figure}

Figure~\ref{fig:2} presents the global Fourier amplitudes for the lowest four modes. The higher-order modes are much smaller and are not shown. The initial episode of disk gravitational instability starts soon after disk formation ($t=34$~kyr) and continues for about 300~kyr. During this time period all four Fourier modes gradually decline. The $m=1$ mode dominated in the initial 150~kyr and the corresponding spiral pattern is irregular and lopsided as can be seen in the upper-left panel of Figure~\ref{fig:1}.
The $m=2$ mode begins to dominate after 200~kyr, manifesting the appearance of a regular two-armed spiral. The second rise in all Fourier amplitudes occurs around 370~kyr, during which the $m=2$ mode dominates. The corresponding two-armed pattern can also be seen in the second row of Figure~\ref{fig:1}. The third episode of disk destabilization takes place around 640~kyr, during which again the $m=2$ mode prevails. Out of all three episodes, the initial disk destabilization soon after disk formation is the strongest. This can be explained by continual mass-loading from the infalling envelope in this phase.


\begin{figure}
\begin{centering}
\includegraphics[width=1\columnwidth]{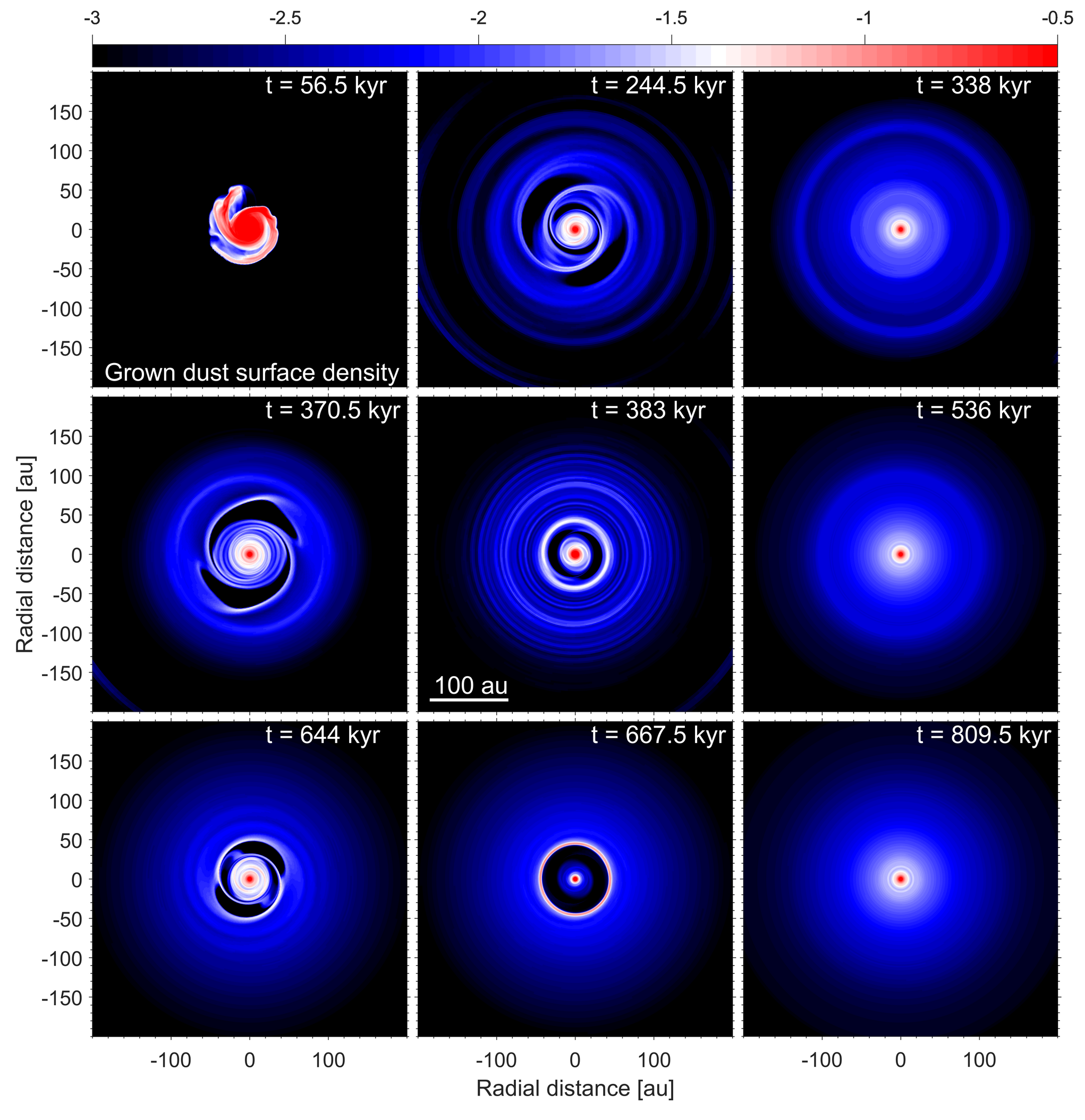}
\par\end{centering}
\caption{\label{fig:1dust} Similar to Fig.~\ref{fig:1} but for the grown dust component. }
\end{figure}

\begin{figure*}
\begin{centering}
\includegraphics[width=2\columnwidth]{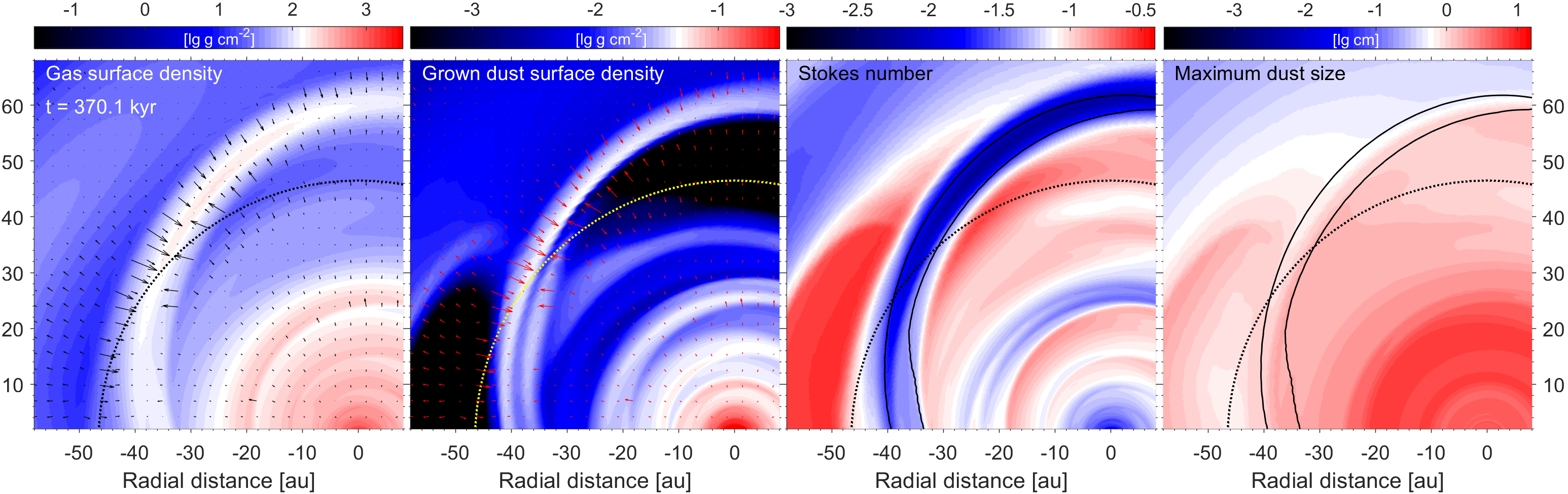}
\par\end{centering}
\caption{\label{fig:10} Drift of grown dust towards the spiral arms. The first and second panels (from left to right) present the residual dust velocity field superimposed on the gas and grown dust surface densities, respectively (in log g cm$^{-2}$). The third and fourth panels show the Stokes numbers and maximum size of dust grains (in log cm), respectively. The dotted circles indicate the position of corotation of the the spiral pattern with the gas disk. The black contour lines outline the position of the spiral for convenience.}
\end{figure*}

Episodic destabilization of a protoplanetary disk was first reported in \citet{2019VorobyovSkliarevskii} and is likely caused by secular redistribution of mass within the disk, resulting in accumulation of matter at tens of astronomical units and ultimately in triggering of gravitational instability.  This phenomenon was found to occur only in models for which the $\xi$-parameter (see Eq.~\ref{Sink:1}) is set to a relatively low value $\le 0.2$. Such a low value of $\xi$ effectively means the presence of a dead zone in the inner 1.0~au. The matter accumulates beyond the dead zone and this leads to recurrent episodes of disk gravitational destabilization. Therefore, the recurrent gravitational instability phenomenon is not expected to occur in disks that lack a persistent dead zone in the innermost disk regions. The initial pre-stellar core parameters ($M_{\rm core}$ and $\beta$) may also affect this phenomenon, as the core mass and angular momentum largely determines the mass of subsequently formed disks \citep{2011Vorobyov} and hence the strength of gravitational instability. The value of $M_{\rm core}=0.59~M_\odot$ is near the peak of the initial core mass function in star-forming regions \citep[e.g.][]{2010Andre} and the value of $\beta=0.24\%$ is within the limits found for dense pre-stellar cores by \citet{2002Caselli}. The chosen model parameters are therefore rather typical for star-forming regions. 
We will return to this phenomenon in a follow-up study. Here, we simply take the initial, second, and third episodes of disk destabilization and consider accretion bursts of various duration and shape to explore their effect on the development and sustainability of gravitational instability in the disk.

Figure~\ref{fig:1dust} presents the spatial distribution of grown dust at the same time instances as for the gas distribution in Figure~\ref{fig:1}.
We consider the distribution of grown dust because it can decouple dynamically from gas and hence this dust fraction is expected to have a spatial pattern that differs from that of gas. On the contrary, the spatial distributions of small sub-micron dust and gas are similar. Indeed, a comparison of Figures~\ref{fig:1} and \ref{fig:1dust} demonstrates that the spatial distribution of grown dust can be different from that of gas. At the time instances when a strong spiral pattern is present in the gas disk (e.g., $t=$244.5, 370.5, and 644~kyr), the spiral arms look much sharper in grown dust than in gas and there exist deep cavities in the inter-arm regions that are strongly depleted in dust. This is the result of grown dust drifting towards local pressure maxima represented by the arms. 

As the spiral pattern in the gas disk diminishes, the dust arms merge into an oval-like structure and finally unto a pronounced ring at several tens of astronomical units ($t$=383 and 676.5~kyr). The ring-like structure in grown dust disappears (or weakens significantly) once the the gas disk stabilizes and attains a nearly axisymmetric state ($t$=536 and 809.5~kyr).
This means that the ring is causally linked to spiral arms, the latter are a manifestation of gravitational instability in the disk. Later we will show that accretion bursts and disk fragmentation can add to the complexity of the ring structure in grown dust. We note that the dust rings in our model are not related to zonal flows \citep[e.g.][]{2019Teague}, since our thin-disk model cannot consider this effect. 

Figure \ref{fig:10} illustrates the process of grown dust drift towards the spiral arms by plotting the residual dust velocities over the gas and grown dust surface density distributions at $t=370.1$~kyr. The residual velocities were calculated by subtracting the gas velocities from those of grown dust.
The Figure also shows the spatial distribution of the Stokes number $St$ and maximum size of grown dust $a_{\rm max}$.
The dotted circles present the position of corotation of the spiral pattern with the gas disk.  Clearly, the grown dust drifts towards the spiral arms, but this process is most pronounced in the vicinity of corotation. The reason for that was studied in detail in \citet{2018VorobyovAkimkin}, who showed that near corotation the characteristic velocities of dust drift are greater than the bulk velocity of dust in the frame of reference of the spiral pattern. Far from corotation, grown dust simply passes with the flow through the spiral arms. 

Inspection of Figure~\ref{fig:10} shows that the contrast in the density between the spiral arm and inter-armed region is much stronger in grown dust than in gas (see also fig.~6 in \citet{2018VorobyovAkimkin}). This means that grown dust drifts out of the inter-armed regions towards the spiral arms, thus increasing the contrast. This process is, however, moderated by lower Stokes numbers as the dust enters the arm. Indeed, the spatial distribution of the Stokes numbers shown in the third panel of Figure~\ref{fig:10} indicates that the strongest spiral arm is distinguished by the lowest values of $St\approx 0.01$, while in the inter-armed region and behind the arm the Stokes numbers can reach 0.3. The Stokes number is directly proportional to the size of grown dust and inversely proportional to the gas volume density and sound speed (hence, square root of temperature). The right panel of Figure~\ref{fig:10} indicates that $a_{\rm max}$ is largest in the innermost disk regions and decreases gradually outwards. There is no strong increase of $a_{\rm max}$ in the vicinity of the arm.
On the other hand, the gas density and temperature both increase in the arm and this acts to decrease the Stokes number there. As a result, the dust is attracted towards the arm but once it enters the arm the drift velocity towards the arm's center decreases. The resulting dust enhancement usually does not exceed a factor of two (see also fig.~6 in \citet{2018VorobyovAkimkin}).

We note that previous works that addressed the phenomenon of dust drift and accumulation of dust in spiral arms considered cases with a spatially constant Stokes number. For instance, \citet{2004RiceLodato} and \citet{2015Dipierro} found that dust particles with $St=1.0$ can strongly concentrate within spiral arms. Our disks, however, are characterized by a spatially varying Stokes number, hardly reaching 0.5 and much smaller in the arms, which hinders the process of dust accumulation. Nevertheless, the spiral arms are clearly seen in the distribution of grown dust before the burst is initiated.

\begin{figure}
\begin{centering}
\includegraphics[width=1\columnwidth]{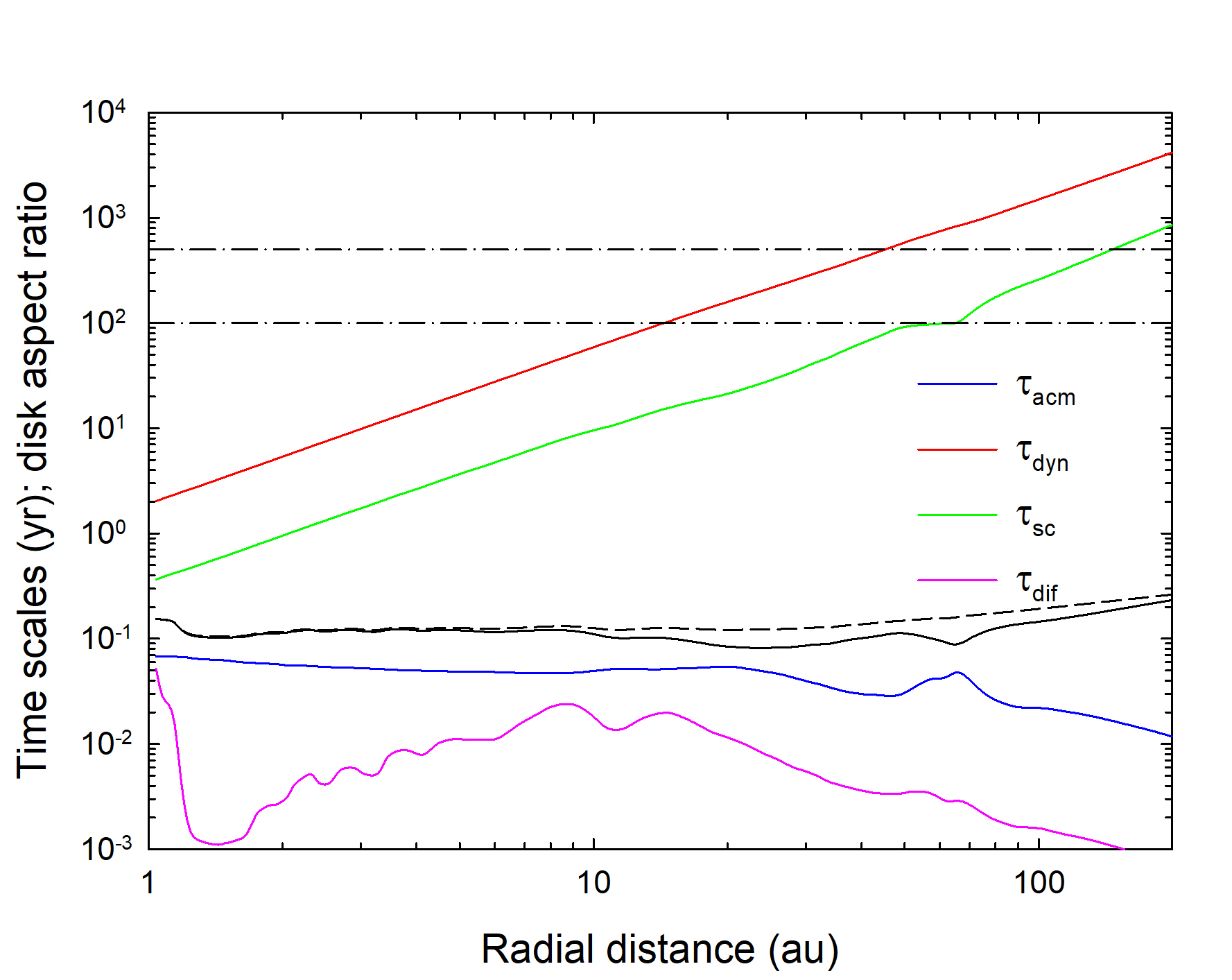}
\par\end{centering}
\caption{\label{fig:1a} Characteristic timescales in the quiescent disk at t=370~kyr. Shown are the dynamical timescale $\tau_{\rm dyn}$ (red line),  vertical equilibrium timescale $\tau_{\rm sc}$ (green line), radiative energy accumulation timescale $\tau_{\rm acm}$   (blue line), and radiative energy diffusion timescale $\tau_{\rm dif}$ (pink line). The black solid and dashed lines show the aspect ratio of the disk (vertical scale height to radial distance) calculated for the model scale height $H$ and Gaussian scale height $H_{\rm G}$, respectively. The dash-dotted lines provide the span of considered burst durations.  }
\end{figure}

\begin{figure}
\begin{centering}
\includegraphics[width=1\columnwidth]{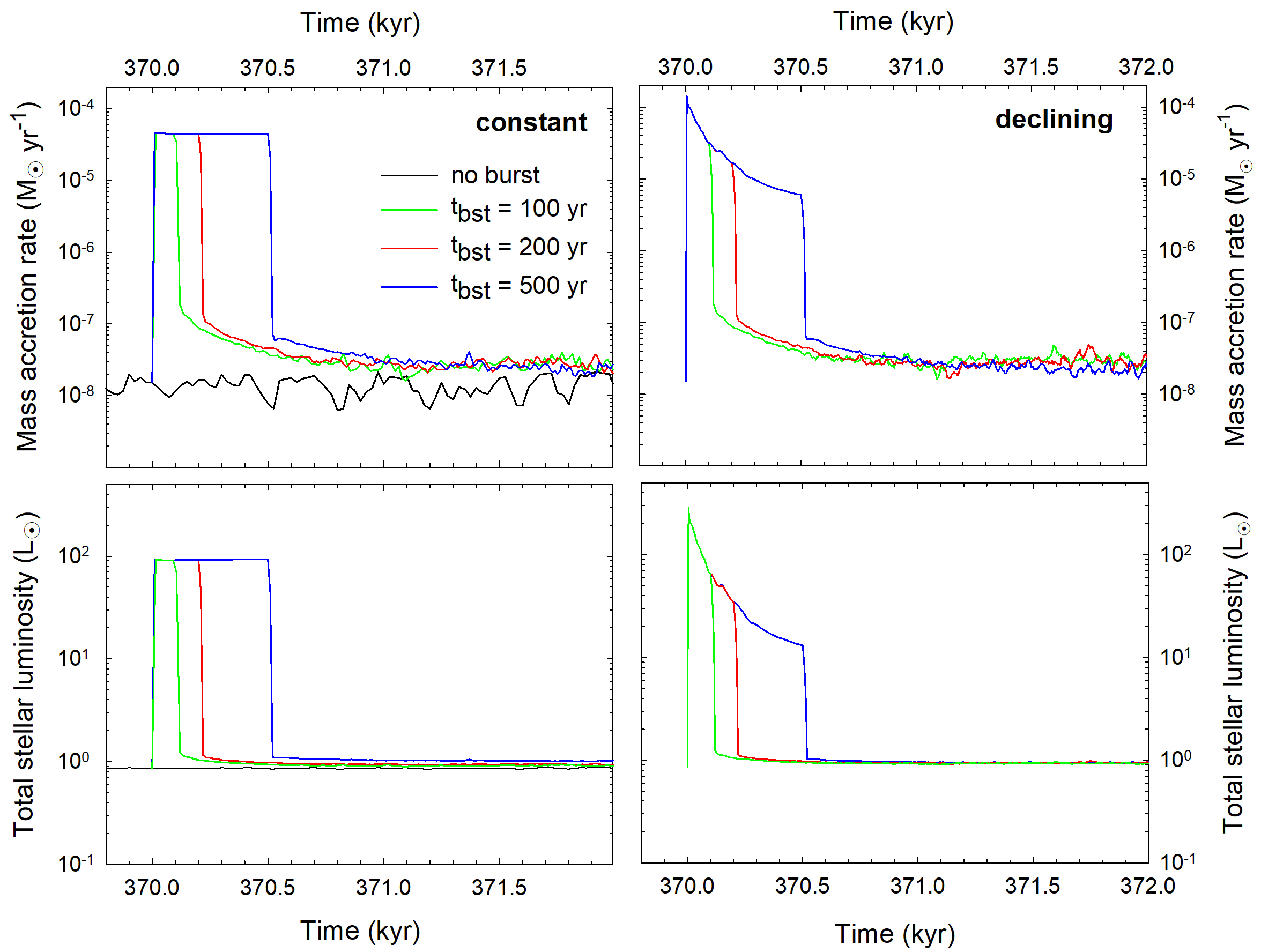}
\par\end{centering}
\caption{\label{fig:2a} Mass accretion (top panel) and total stellar luminosity (bottom panel) versus time for the cases without the burst and with the bursts of various duration. The left column corresponds to the bursts with a constant magnitude, while the right columns shows the bursts declining with time.}
\end{figure}

\section{Response of the disk to accretion bursts}
\label{short_burst}

\subsection{Characteristic timescales}
Before initiating the burst it is worthwhile to consider characteristic timescales of the system, which can be useful when analysing the response of the disk to the burst. These are
the dynamical timescale $\tau_{\rm dyn}=2\pi\Omega^{-1}$, where $\Omega$ is the local angular velocity that takes into account the gravity of the star and enclosed disk, the timescale for diffusion of stellar radiation to the the disk midplane $\tau_{\rm dif}$, the timescale for accumulation of radiation energy $\tau_{\rm acm}$, and the timescale for the propagation of sound waves in the vertical direction
$\tau_{\rm s}$. The diffusion timescale is defined as
\begin{equation}
    \tau_{\rm dif} =  {H^2 \over D_{\rm rad}}= \kappa_{\rm R} \Sigma_{\rm d,tot} {H \over \lambda c},
\end{equation}
where $c$ is the speed of light, $D_{\rm rad}$ is the diffusion coefficient of stellar radiation, and $\lambda$ is the flux limiter, which approaches 1/3 in the optically thick limit. The energy accumulation timescale can be defined as
\begin{equation}
    \tau_{\rm acm} = {e \over F_{\rm irr}}.
\end{equation}
This timescale describes how rapidly the temperature
of the medium can grow when capturing the radiative energy. The timescale $\tau_{\rm sc}$ is defined as
\begin{equation}
 \tau_{\rm sc} = {H \over c_{\rm s}  }
\end{equation}
and describes how quickly the disk can attain a vertical equilibrium.

Figure~\ref{fig:1a} presents the corresponding timescale at $t=370$~kyr calculated from the azimuthally-averaged quantities. The vertical dash-dotted lines indicate the span of burst durations considered in this work. Clearly, the thermal timescales $\tau_{\rm dif}$ and $\tau_{\rm acm}$ are much shorter than the dynamical and vertical timescales ($\tau_{\rm dyn}$ and $\tau_{\rm sc}$) and also much shorter than the burst durations, meaning that the disk thermal balance can quickly adjust to the burst. The vertical timescale $\tau_{\rm sc}$ is shorter than the dynamical timescale and is shorter than the burst duration inside 70--180~au, meaning that the vertical disk structure can also adjust reasonably fast to the burst, at least in the region of interest where the spiral structure is localized. 
On the other hand, the dynamical timescale $\tau_{\rm dyn}$ is shorter than the burst only inside 10--45~au, depending on the burst duration, while the spiral structure extends to 60--70~au. We also calculated the revolution period of the spiral pattern and it turned out to be comparable (560 yr) to the longest burst duration of 500~yr.
As similar trend is found for other considered evolutionary times and for the data taken along individual radial cuts (not azimuthally averaged).

Finally, we note that the ratio of disk vertical height to radial distance is much smaller than unity (see the solid black line in Fig.~\ref{fig:1a}), justifying the thin-disk approximation adopted in our modeling.  The vertical scale height of our disk is computed taking the disk self-gravity into account \citep{VorobyovBasu2009}. We compared the resulting values of $H$ with the Gaussian scale height $H_{\rm G}=\sqrt{c_s^2 r^3/(GM_\ast)}$ in the limit of negligible disk self-gravity and found that $H<H_{\rm G}$ only beyond 10~au (see the dashed black line). The ratio $H/H_{\rm G}$ varies in the 0.55--1.0 limits throughout the disk, with the minimum value being in good agreement with the shearing-box 3D numerical simulations of \citet{2017RiolsLatter}.
The values of $H$ in our models are also in agreement with the 2+1D numerical simulations that perform the reconstruction of the disk vertical density and thermal structure \citep{2017VorobyovPavlyuchenkov}. We also note that in the limit of negligible disk self-gravity the vertical timescale is  $\tau_{\rm sc}=\Omega^{-1}$ and it is smaller than the dynamical timescale $\tau_{\rm dyn}$ by a factor of $2\pi$. A similar relation holds also for self-gravitating disks with the corresponding factor lying in the 4.8--8.2 limits (see the red and green lines in Fig.~\ref{fig:1a}).

\begin{figure}
\begin{centering}
\includegraphics[width=1\columnwidth]{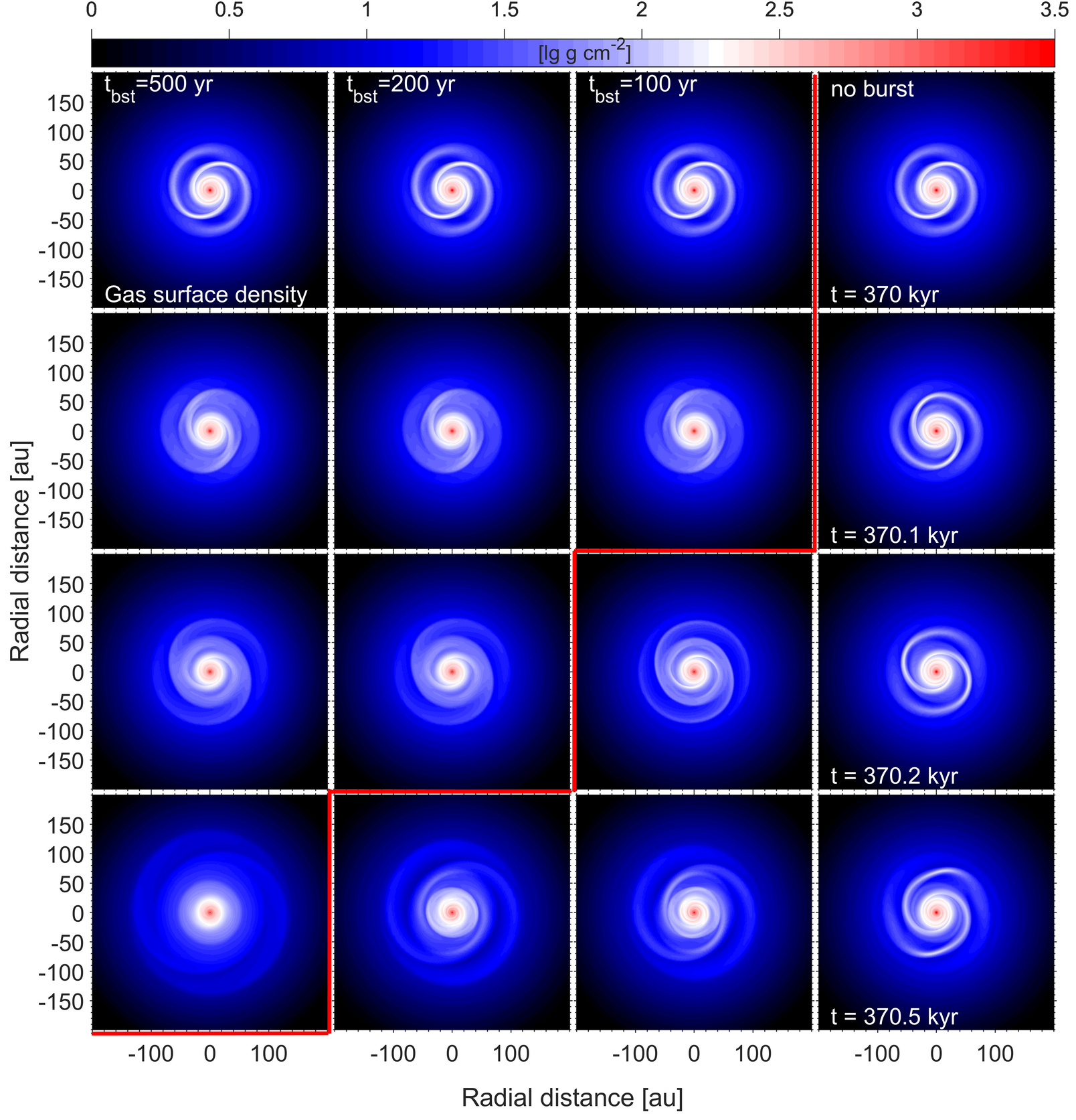}
\par\end{centering}
\caption{\label{fig:3} Response of the spiral pattern to bursts of different duration. Shown is the gas surface density in the inner $400\times 400$~au$^{2}$ box. Columns from left to right represent models with different burst durations: 500~yr, 200~yr, 100~yr, and the model without the burst. The initial 500~yr since the onset of the burst with a constant-magnitude shape are shown. The red line separates disk images in the burst phase from those in the post-burst or no-burst phase. The scale bar is in log  g~cm$^{-2}$. }
\end{figure}

\begin{figure}
\begin{centering}
\includegraphics[width=1\columnwidth]{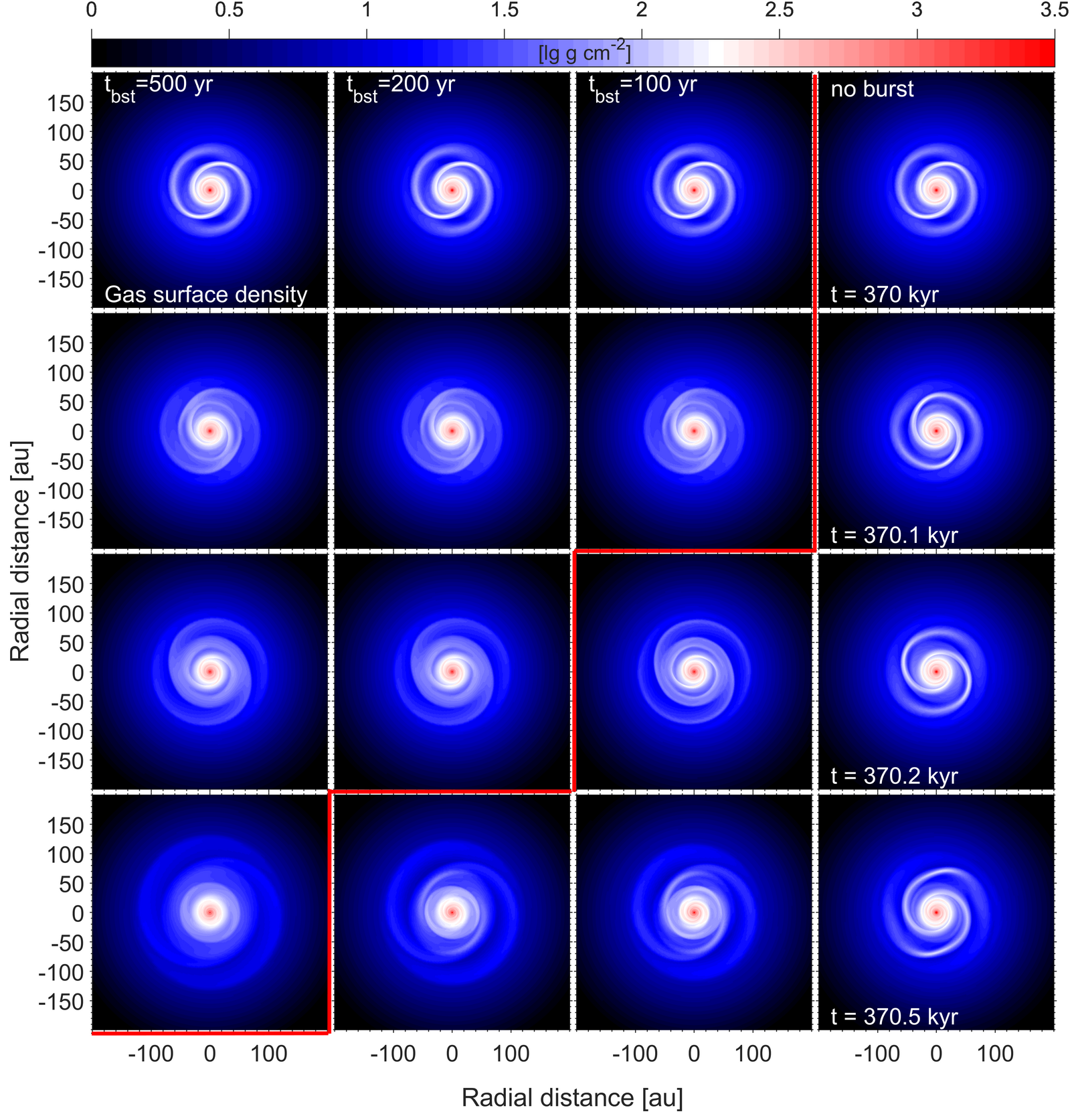}
\par\end{centering}
\caption{\label{fig:3a} Similar to Figure~\ref{fig:3}, but for the burst with a time-declining magnitude.}
\end{figure}

\subsection{Characteristics of the burst}
We have chosen three time instances during the entire disk evolution that are characterized by a well-developed two-armed spiral structure: $t=245$~kyr, $t=370$~kyr, and $t=645$~kyr. To initiate an accretion burst, we first 
note that the gas temperature at the CSC--disk interface is greater than  1300~K at these evolutionary times. This value is sufficient for thermal ionization of alkaline metals  and ignition of the MRI \citep{2015Desch}. We therefore restart our numerical simulations at the corresponding time instances with $T_{\rm crit}$ set equal to 1300~K. Note that at earlier times $T_{\rm crit}$ was set to 1600~K (see Sect.~\ref{SCS}) and hence the system was in a quiescent phase. Lowering the critical temperature, we immediately trigger the MRI burst. In this sense, our simulations are not fully self-consistent, but it is not well known at what temperatures the MRI ignition takes place, and there exists a certain window in the values of $T_{\rm crit}$ that trigger the MRI. The time period of the burst was manually set equal to 100, 200, and 500~yr. These burst durations are consistent with what was found in numerical hydrodynamics simulations of MRI-triggered bursts \citep{2014BaeHartmann,2014AudardAbraham}.  We found that longer bursts ({\bf$ > 500$~yr})  deplete the CSC of matter (the gas density goes to a negative value) and are therefore not feasible for our model realization.

The shape of the burst was chosen to follow two patterns: constant with time and gradually declining with time. These patterns reflect two typical light curves observed in FU~Orionis and V1057 \citep[see, e.g., fig. 3 in][]{1996HartmannKenyon}, although deviations from these two limiting cases are possible \citep{2014AudardAbraham}.
In the first case,  $\dot{M}_{\rm \ast,bst}$ (see~Eqs.~\ref{csc1} and \ref{csc2}) is set equal to $5\times10^{-5}~M_\odot$~yr$^{-1}$ during the entire burst duration. In the second case, the mass accretion rate has the following form
\begin{equation}
    \dot{M}_{\rm \ast,bst}(t) = \left( \dot{M}_{\rm \ast,max} \over \Sigma_{\rm csc,0} \right) \Sigma_{\rm disk}(t),
\end{equation}
where $\dot{M}_{\rm \ast,max}$ is the maximum accretion rate at the onset of the burst, set equal to $2\times10^{-4}~M_\odot$~yr$^{-1}$, $\Sigma_{\rm csc,0}$ is the gas surface density in the CSC at the onset of the burst, and $\Sigma_{\rm disk}(t)$ is the azimuthally averaged gas surface density in the disk layer immediately adjacent to the CSC. While the quantity in the parentheses stays constant during the burst, $\Sigma_{\rm disk}(t)$ gradually declines as the burst develops and drains the matter in the innermost disk regions. As a result, the burst magnitude also declines with time. For both shapes of the accretion burst, $\dot{M}_{\rm \ast,bst}$ is re-set back to zero once the burst is over.

Figure~\ref{fig:2a} shows the corresponding mass accretion rates and total (accretion plus stellar photospheric) luminosities as a function of time. In the constant-magnitude case, the luminosity increases by about a factor of 100 during the burst and stays almost constant at around $1.0~L_\odot$ after the end of the burst in the considered evolution period. In the time-declining case, the peak luminosity of about $300~L_\odot$ is followed by a gradual decline to $15~L_\odot$ until the burst is over. These values are within the limits measured for FUors \citep{2014AudardAbraham}. The mass accretion rate in the post-burst epoch shows only small-scale flickering due to perturbations in the flow caused by the global gravitational instability of the disk.

\begin{figure}
\begin{centering}
\includegraphics[width=1\columnwidth]{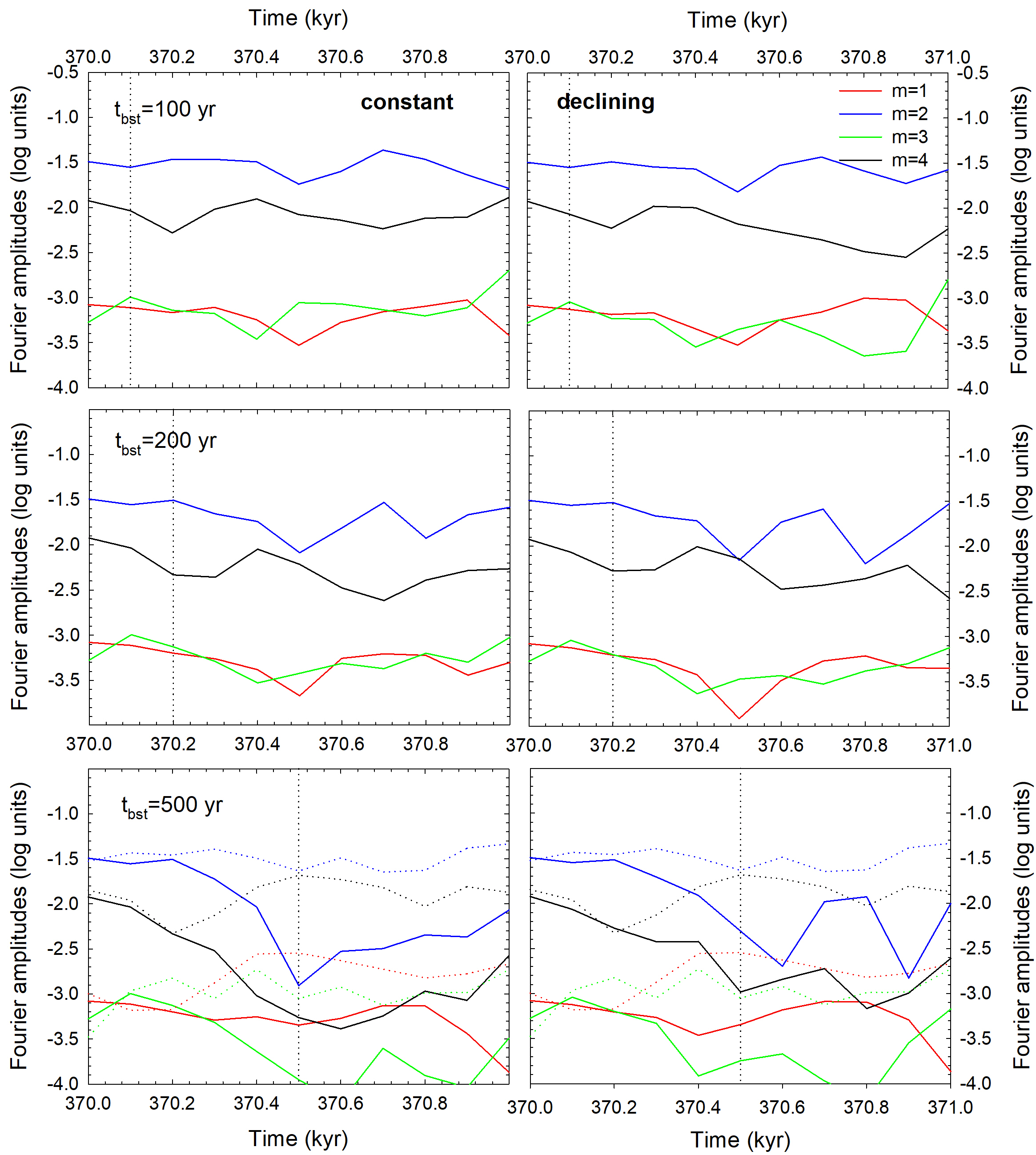}
\par\end{centering}
\caption{\label{fig:3b} Global Fourier amplitudes vs. time in models with different burst durations $t_{\rm bst}$. The four lowest modes $m=$1, 2, 3, and 4 are shown. The left and right columns correspond to the constant-magnitude and declining-magnitude bursts, respectively. The vertical dotted lines indicate the end of the burst in each model. The dotted lines in the bottom row show Fourier amplitudes for the model without the burst for comparison.}
\end{figure}

\subsection{Response of the gas disk to the burst}
We now proceed with analysing the response of the disk to accretion bursts. Figures~\ref{fig:3} and \ref{fig:3a} present the evolution of the gas disk in models with bursts of various duration and shape.  In particular, Figure~\ref{fig:3} correspond to the constant-magnitude burst, while Figure~\ref{fig:3a} shows the results for the burst magnitude declining with time. The columns from left to right show the models with $t_{\rm bst}$=500, 200, and 100~yr. The last column on the right hand side presents the case without the burst for comparison.  Bursts in all models are triggered at the same time $t=370$~kyr, but only the initial 500~yr of disk evolution are shown. The red contour line separates the evolutionary instances undergoing the burst from those in the post-burst or no-burst state. 

Clearly, bursts of any considered duration influence the shape of the spiral structure in the gas disk, but the effect  is weakest for the shortest burst with $t_{\rm bst}=100$~yr and is strongest for the longest burst with $t_{\rm bst}=500$~yr. In this model, the sharp spiral pattern completely diminishes by the end of the burst, leaving behind only tracers of weak spiral arms around 100~au. In the time-declining burst model, a weak ring-like structure forms in addition at several tens of astronomical units. In the shortest burst model with $t_{\rm bst}=100$~yr, the spiral structure loses its integrity and sharpness by the end of the burst, but is still discernible throughout the subsequent 400~yr of evolution. A ring-like structure also forms in this case and the spiral arms seem to emanate from the ring. 

\begin{figure*}
\begin{centering}
\includegraphics[width=2\columnwidth]{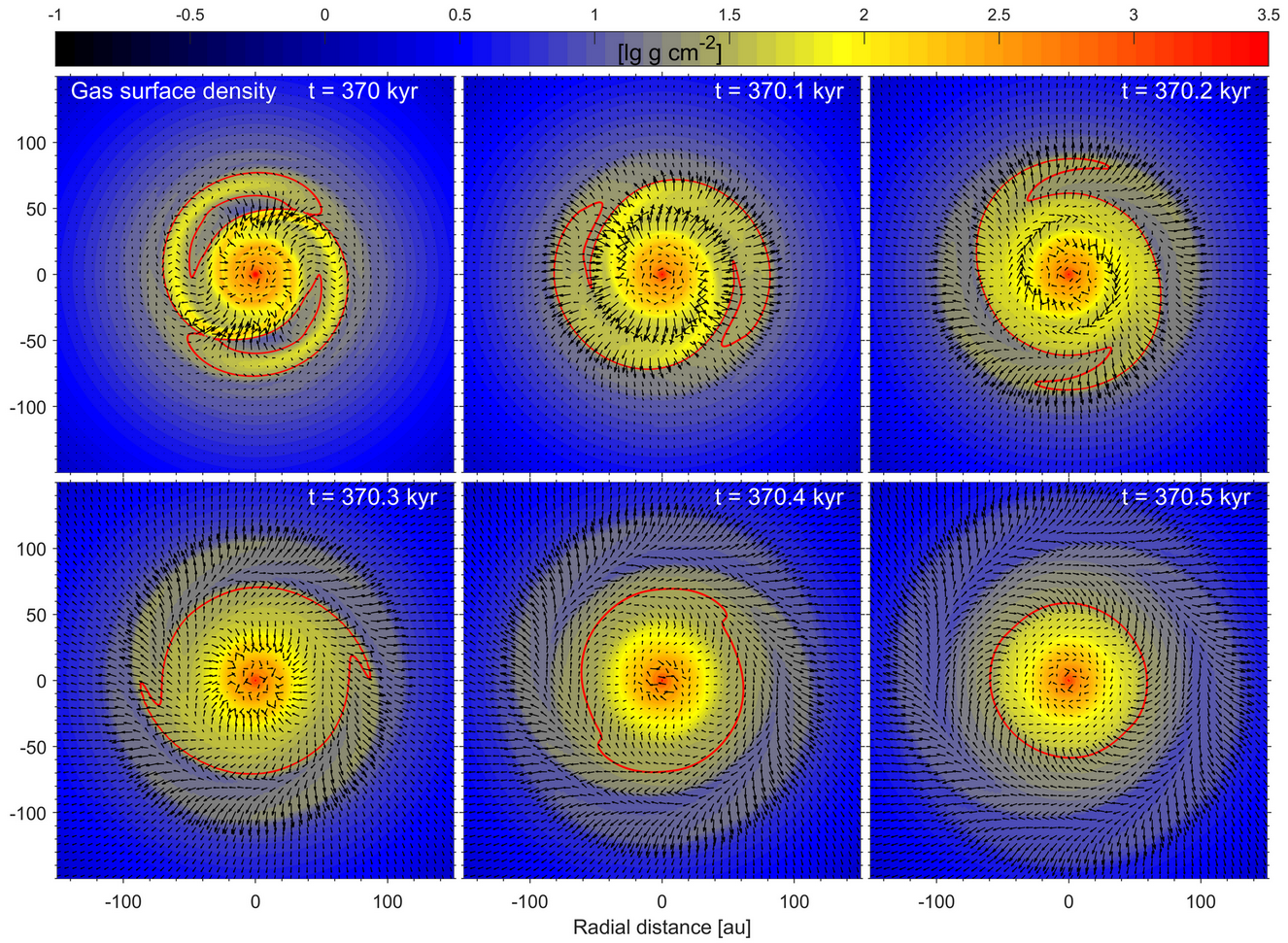}
\par\end{centering}
\caption{\label{fig:11} Residual gas velocity field superimposed on the gas surface density. Six time instances are shown for a constant luminosity burst of 500~yr in duration. The burst is triggered at 370~kyr. The residual velocity field is obtained by subtracting the Keplerian rotation. The red contour lines are the isodensities with $\Sigma_{\rm g}$=30~g~cm$^{-2}$.}
\end{figure*}

We check the other two time instances for the onset of the burst at $t=245$~kyr and $t=645$~kyr and confirm the general trend that the longest-burst model ($t_{\rm bst}$=500~yr) destroys the spiral pattern and turns the gas disk in almost an axisymmetric state by the end of the burst. The shortest-burst model, on the other hand, can only weaken and diffuse the spiral pattern, which is still visible by the end of the burst. The shortest-burst model also features a ring-like structure that forms at several tens of astronomical units 400~yr after the burst ends.

The response of the disk to luminosity bursts can be understood with the help of characteristic timescales presented in Figure~\ref{fig:1a} and residual gas velocities shown in Figure~\ref{fig:11} for the $t_{\rm bst}=500$~yr burst. The residual velocities were calculated by subtracting the regular Keplerian rotation from gas velocities, taking both the stellar and enclosed disk masses into account when calculating the Keplerian velocity. The 500-yr-long burst is shorter than the dynamical timescale inside 45~au and is comparable to the revolution period of the spiral but is much longer than the thermal timescales. As a result, the disk is almost instantaneously heated by the burst. The gas pressure rises and drives the disk out of equilibrium, forcing it to expand. This expansion motion is clearly seen in Figure~\ref{fig:11}. The spiral pattern smears out almost entirely due to expansion by the end of the burst ($t=370.5$~kyr).  The bursts of shorter duration (100 and 200~yr) also heat up the disk because the thermal timescales are still much shorter,  but the burst duration is too short compared to the dynamical timescale and revolution period of the spiral pattern. As a result, the spiral pattern somewhat diffuses due to expansion but does not disappear entirely.

For a more quantitative analysis of the effects of the burst on the disk spiral pattern we plot in Figure~\ref{fig:3b} the global Fourier amplitudes defined by Equation~(\ref{fourier}). A time period of 1000~yr is shown starting from the onset of the burst, with the vertical dashed lines indicating the considered burst durations. The lowest four modes are analyzed for the constant-magnitude burst (left column) and declining-magnitude burst (right column). The bottom row also shows the Fourier amplitudes for the burstless model for comparison.  Clearly, the $t_{\rm bst}=100$~yr and $t_{\rm bst}=200$~yr bursts do not affect notably the Fourier modes. A small decline is noticeable, but the effect is incomparable to what happens in the $t_{\rm bst}=500$~yr case -- the Fourier amplitudes drop by 1.0--1.5~dex  by the end of the burst. The spiral structure does not re-generate to its pre-burst  state for the next 500~yr.

\begin{figure}
\begin{centering}
\includegraphics[width=1\columnwidth]{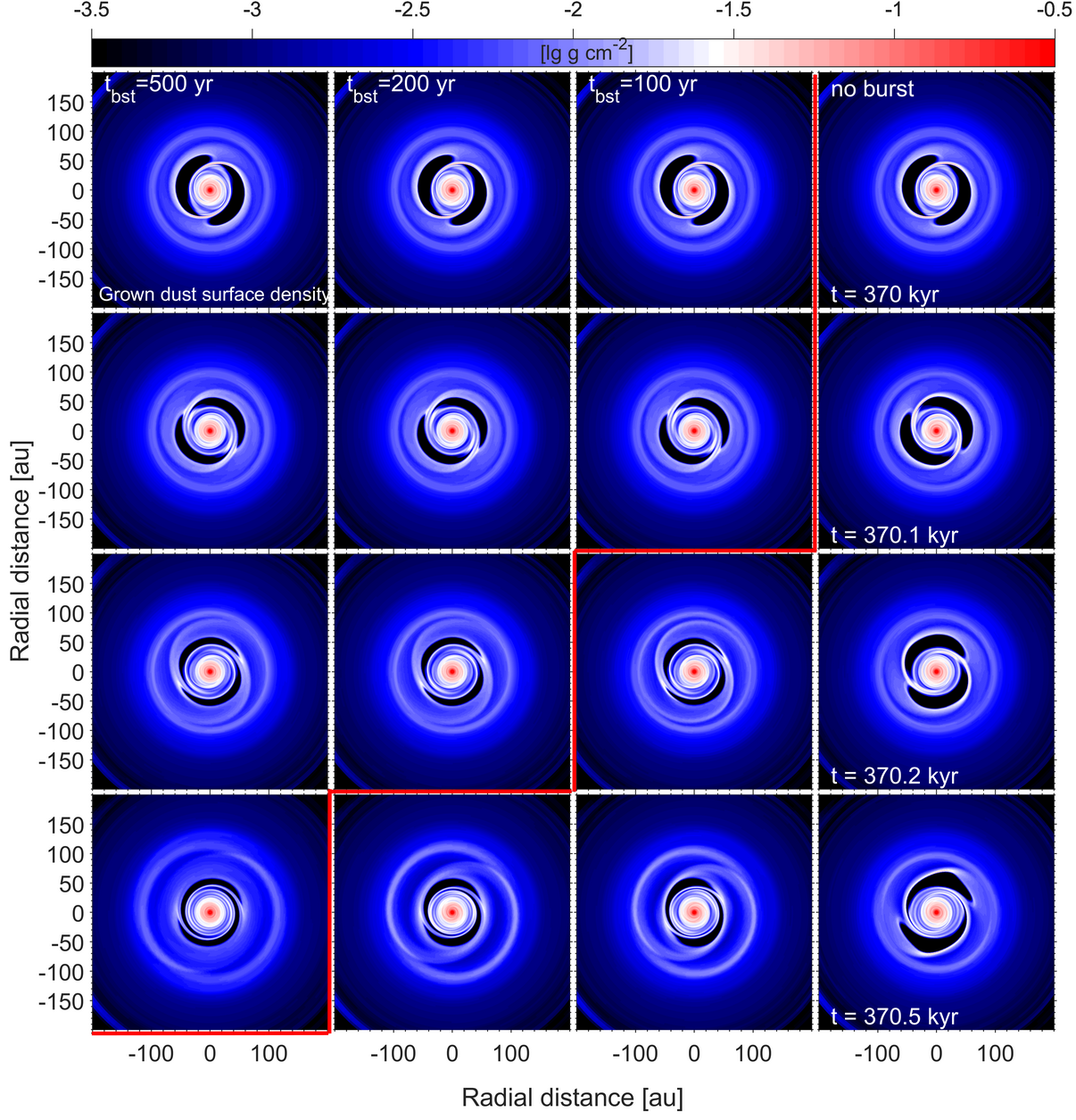}
\par\end{centering}
\caption{\label{fig:4} Similar to Figure~\ref{fig:3}, but for the grown dust component.}
\end{figure}

\subsection{Response of the dust disk to the burst}
Here we consider the spatial distribution of grown dust in the models with different burst durations. Figure~\ref{fig:4} presents the surface density maps of grown dust in the inner $400\times400$~au$^2$ box for the same models, burst durations, and burst shape (constant-magnitude) as in Figure~\ref{fig:3}.
By the end of the longest-duration burst ($t_{\rm bst}=500$~yr) the dust spiral arms are almost gone, but the cavities turn into a deep gap at a radial distance of 40~au. The depth of the gap reaches two orders of magnitude compared to
the grown dust surface density on both sides of the gap. A weak outer ring is also present. The shorter-duration bursts also affect the distribution of grown dust  but its spatial distribution shows a complex morphology with cavities and arms, which can be considered as a transient phase between the spiral-dominated and ring-dominated state.

The effect of the time-declining burst is similar to that of a time-constant burst and is not shown here for the sake of saving space. We also checked the other burst onset time instances (245~kyr and 645~kyr) and confirmed the general trend described above. We conclude that the general outcome of the burst can be described as a transformation of the spiral-like initial distribution with deep cavities into a ring-like distribution with deep gaps. This trend is best expressed for the bursts of longest duration on the order 500~yr.  The transformation of a spiral-like distribution of grown dust to a ring-like distribution was already found in the context of global disk evolution in Figure~\ref{fig:1dust}. There we noted that the transformation occurs when the spiral pattern in the gas disk gradually diminishes as the disk stabilizes after a recurrent episode of gravitational instability. Here the disk stabilization occurs due to burst heating followed by expansion (see Fig.~\ref{fig:11}).  

The formation of ring-like structures in our models is intriguing. The presence of rings is often taken as evidence in favour of planet formation \citep[e.g.,][]{2015DongZhu}. In our case, however, no planets are formed, and the rings occur purely due to thermo-hydrodynamic effects of the burst related to dust and gas friction and decoupling between gas and dust subsystems. In a follow-up study we plan to explore the detectability of these ring-like structures with submillimeter interferometers, such as ALMA. 

In the burst context, the case of V883~Ori is of particular interest. This system is a FUori-like object, according to classification of \citet{2014AudardAbraham}, meaning that the onset of the burst has not been documented. 
Note that V883~Ori has a rather high disk mass of $\ga0.3~M_\odot$ that is sufficient for the development of GI  \citep{2013Vorobyov}. Moreover, V883~Ori does not have an obvious companion or a planet, suggesting that the burst must have been caused either by the MRI that is assisted by gravitational instability to ignite the burst \citep{2014BaeHartmann} or infall of gaseous clumps formed via disk fragmentation \citep{VorobyovBasu2015}. In both cases, the disk must have been unstable before the burst.
According to our simulations, a burst of 100--200~yr in duration cannot fully destroy the spiral structure in the gas disk. One needs a longer burst of several hundred years in duration. V883~Ori may already be more than 100 yr in outburst \citep{1890Pickering} and at this stage we may expect a notable weakening of the spiral structure that presumably existed in the pre-burst gas disk. Concurrently, the dust disk is expected to show a ring-like morphology with gaps. Although the parameters of our model are different from those of V883~Ori, the principal effect of the burst on the disk morphology is not expected to be qualitatively different.

Unfortunately, we lack high-resolution observations of the gas disk of V883~Ori that may give us a hint on the absence or presence of strong spirals in the gas component. On the other hand, the spatially resolved image in the millimeter emission (which traces grown dust) does show signatures of rings and gaps, but these structures were interpreted as the result of a shifted snow line of water and increased opacity depth interior to the snow line \citep{2016Cieza}. Our model demonstrates that there can be other explanations and  ring-like structures can form naturally in disks of outbursting stars.


\begin{figure}
\begin{centering}
\includegraphics[width=1\columnwidth]{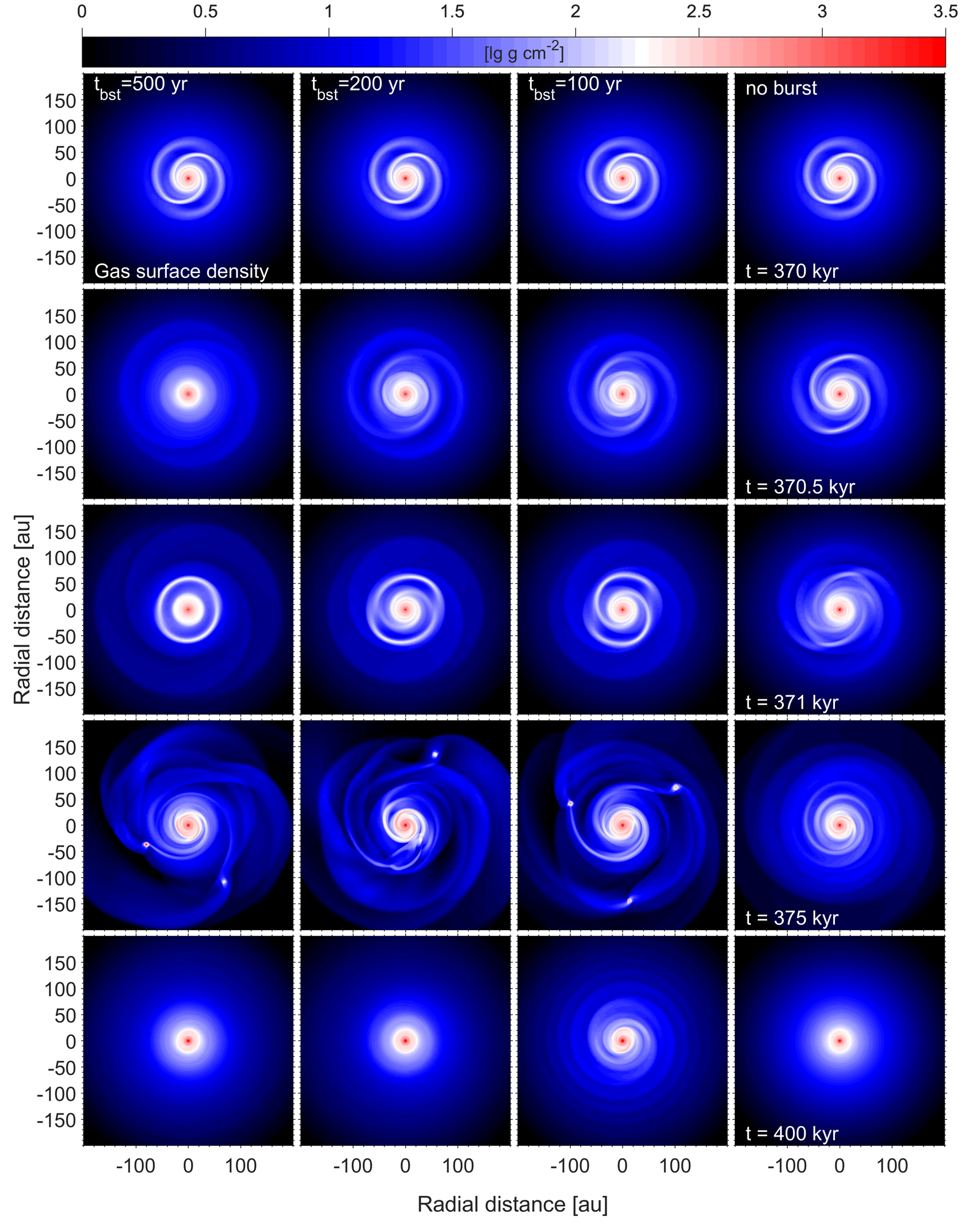}
\par\end{centering}
\caption{\label{fig:5} Gas disk evolution in the model without the burst (right-hand-side column) and in the models with the bursts of various duration (from left to right): $t_{\rm bst}$=500~yr -- first column, $t_{\rm bst}$=200~yr -- second column, $t_{\rm bst}$=100~yr -- third column. Shown are four time instances starting from the onset of the burst: 0.5~kyr (second row), 1.0~kyr (third row), 5.0~kyr (fourth row), and 30~kyr (bottom row).  The scale bar shows the gas surface density in g~cm$^{-2}$ (log units). }
\end{figure}

\section{The long-term effects of the burst}
\label{long_burst}
Now we turn to the long-lasting effects (several tens of kyr) of luminosity bursts on the structure and evolution of the gas-dust disk. First, we consider the burst at $t=370$~kyr and then we discuss the results for the other two burst onset times. Figure~\ref{fig:5} shows the gas disk evolution in models with and without accretion bursts.  Each row presents a sequence of disk images at different times after the onset of the burst spanning a range from 0.5~kyr to 30~kyr. The columns from left to right correspond to models of decreasing burst duration. The last column on the right-hand side shows the model without the burst.

Clearly, the burst has a profound long-term effect on the disk evolution, in accordance with what was also found in \citet{2011Stamatellos}. In the no-burst model, the spiral structure gradually diminishes with time and finally the disk acquires an axisymmetric form at 400~kyr. In the burst models, the initial episode of disk stabilization is shortly followed by vigorous disk destabilization and fragmentation. The fragmenting state does not last long and by $t=400$~kyr the disk turns into an axisymmetric form, except for the $t_{\rm bst}$=100~yr model for which the spiral pattern is still visible.

\begin{figure}
\begin{centering}
\includegraphics[width=1\columnwidth]{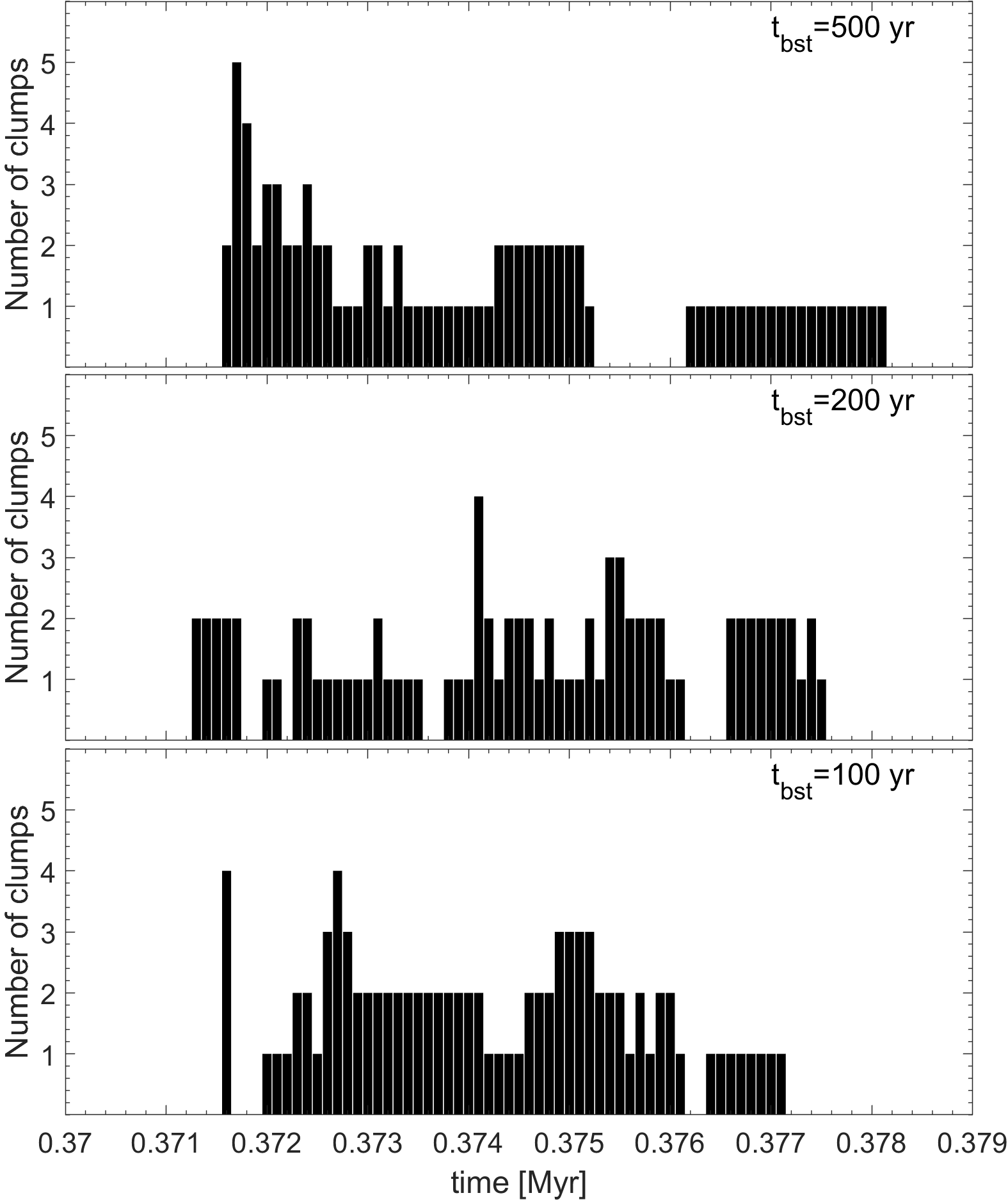}
\par\end{centering}
\caption{\label{fig:6} Number of clumps present in the disk versus time in models with various duration of the burst. }
\end{figure}

It is interesting to review the properties of the clumps formed as a result of the burst. Figure~\ref{fig:6} presents the number of clumps in the disk as a function of time. We used the clump tracking algorithm presented in \citet{2013Vorobyov} to search for the gravitationally bound and pressure-supported clumps in the disk every 100~yr (the frequency of our data output). Disk fragmentation starts about 1.2--1.5~kyr after the burst in the models with $t_{\rm bst}\le 500$~yr. The number of clumps that is simultaneously present in the disk varies from one to five, but most of the time only one or two clumps are present. Clumps disappear 7--8~kyr after the burst. They dissolve through the action of tidal torques as they migrate inward. 

\begin{figure}
\begin{centering}
\includegraphics[width=1\columnwidth]{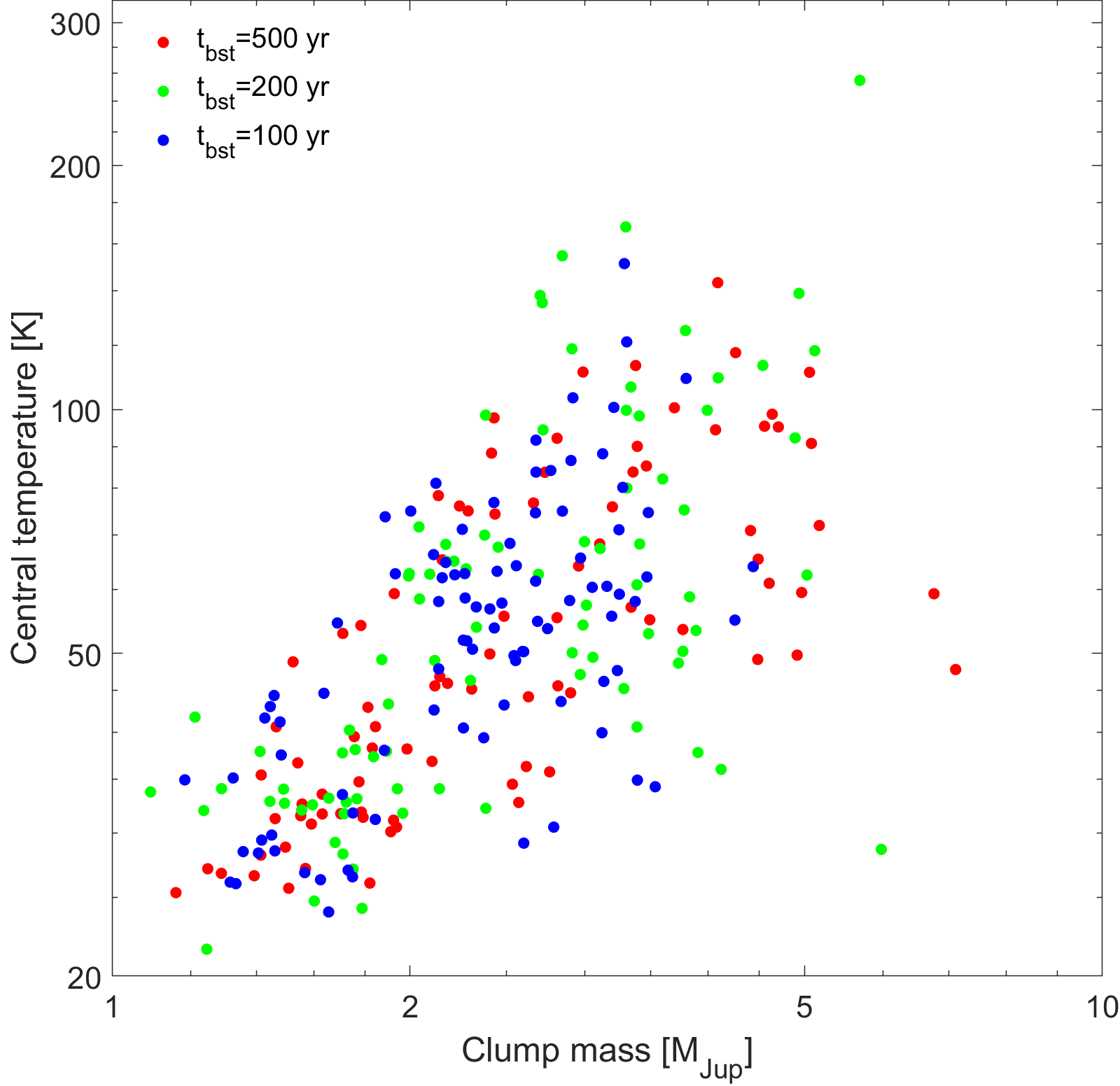}
\par\end{centering}
\caption{\label{fig:6a} Clump central temperature versus clump mass in models with various duration of the burst as shown in the legend.  }
\end{figure}

Figure~\ref{fig:6a} presents the clump central temperature versus clump mass diagram for the three models with disk fragmentation. The figure includes the clumps that are identified every 100~yr of disk evolution using the clump-tracking algorithm, meaning that some long-living clumps may be shown several times, but at different evolutionary time instances. The clump masses vary from one to seven Jupiter masses and the maximum  temperature in the clump interiors reaches 250~K. This means that disk fragmentation forms protoplanetary embryos, rather than brown-dwarf embryos. However, the gas temperature in the interiors of these clumps is too low to initiate second collapse through molecular hydrogen dissociation and form protoplanetary cores.  We should note here that our numerical resolution is sufficient to resolve the Jeans length by more than four grid zones and fulfill the Trulove criterion \citep{1998Truelove}. For instance, the size of the grid cell at 100~au is about 1.7~au and the Jeans length defined as \citep{2013Vorobyov}
\begin{equation}
R_{\rm J} = {c_{\rm s}^2 \over \pi G \Sigma_{\rm g}}
\end{equation}
is approximately 20~au, where $c_{\rm s}$ is the sound speed and $G$ is the gravitational constant. 
On the other hand, the resolution is
insufficient to study the internal structure of the clumps, and the internal temperature may increase with better resolution. The burst-triggered fragmentation needs to be further studied with a higher numerical resolution and for a wider range of disk parameters to make firm conclusions about the feasibility of giant planet formation via this mechanism. 

\begin{figure}
\begin{centering}
\includegraphics[width=1\columnwidth]{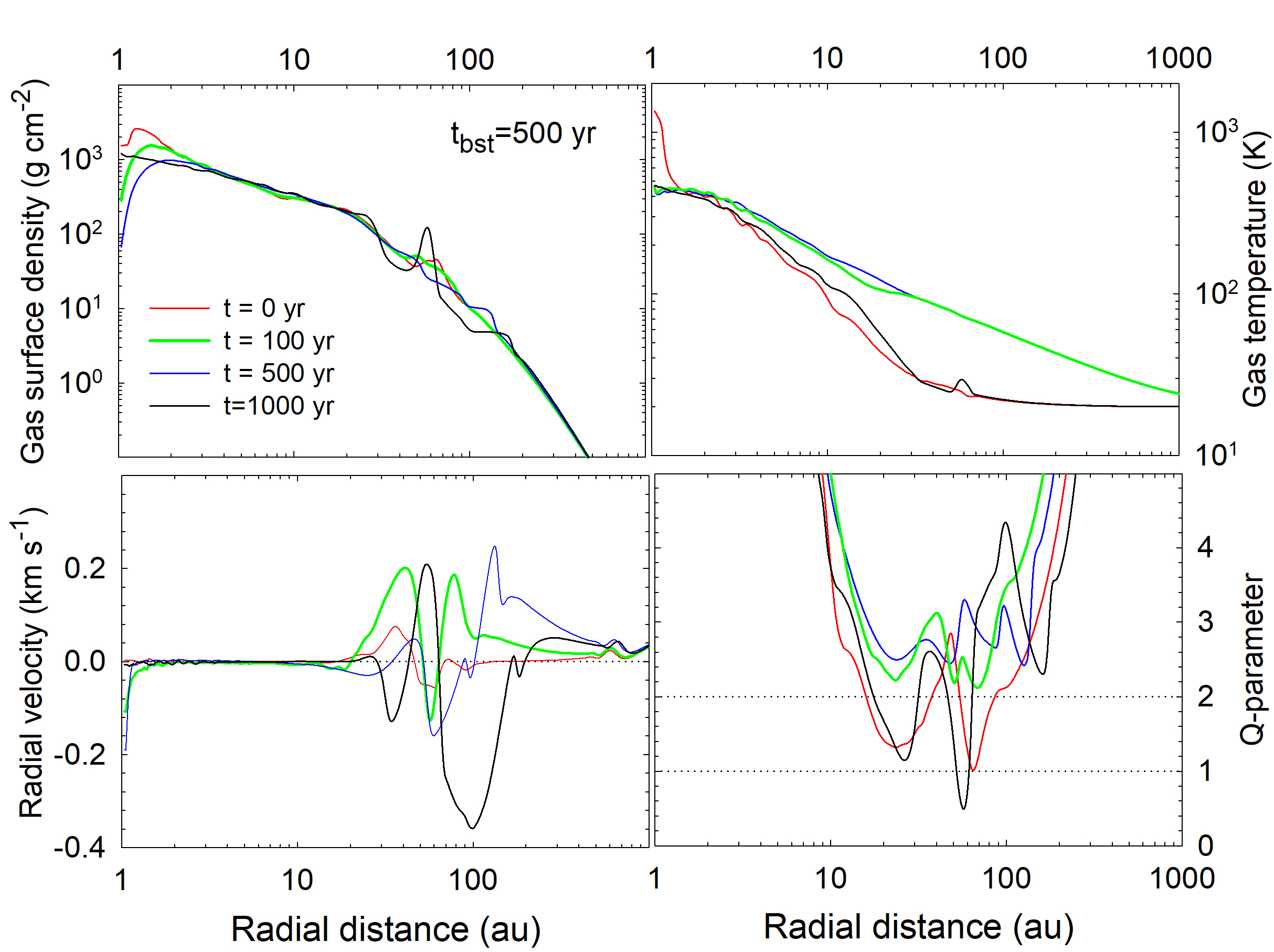}
\par\end{centering}
\caption{\label{fig:7} Azimuthally averaged radial profiles of the gas surface density (top-left), gas temperature (top-right), gas radial velocity (bottom-left), and Toomre $Q$-parameter (bottom-right) at three times after the onset of the burst: 200~yr (green lines), 500~yr (blue lines), and 1000~yr (black lines). The initial profiles are also shown with the red lines. The burst duration is 500~yr with a constant magnitude. The horizontal dotted lines mark zero radial velocity and Q=1 for convenience.   }
\end{figure}

The effect of accretion bursts can be understood if we consider the radial profiles of gas surface density, temperature, radial velocity of gas, and $Q$-parameter shown in Figure~\ref{fig:7} at different time instances after the onset of the burst. For that we have chosen the $t_{\rm bst}=500$~yr case with a constant magnitude of the burst. The initial profiles at the burst onset are also shown with the red lines. Initially, the inner disk regions are characterized by high temperatures and gas densities that are proper for the MRI ignition. At the end of the burst (blue lines), the innermost disk region is drained of matter, but the gas surface density in the rest of the disk is largely unaffected.  The burst evacuates about $25~M_{\rm Jup}$ of the gas disk mass, which is more than 10\% of the entire disk mass, but the effect is localized to the innermost several astronomical units. The effect of the burst on the thermal structure is however global. The temperature notably increases throughout most of the disk extent compared to the pre-burst state. In the post-burst phase, the gas surface densities and temperatures return to the pre-burst values everywhere except for the innermost disk region and also for the outer disk region where disk gravitational fragmentation sets in.  The temperature increase during the burst and its decrease after the burst are very fast due to short thermal timescales in the disk  (see Fig.~\ref{fig:1a}). The increased temperature also raises the gas pressure and drives the disk out of equilibrium. 

Now let us consider the radial velocities of gas and $Q$-parameter shown in the bottom row of Figure~\ref{fig:7}. The Toomre $Q$-parameter is calculated using the upgraded formula that takes dust into account \citep{2018VorobyovAkimkin}
\begin{equation}
Q={c_{\rm d} \Omega \over \pi G (\Sigma_{\rm g}+\Sigma_{\rm d,tot})},
\label{ToomreQ}
\end{equation}
where $\Sigma_{\rm d,tot}$ is the total dust surface density, $\Omega$ the local angular velocity,
$c_{\rm d}=c_{\rm s}/\sqrt{1+\xi}$ the modified sound speed 
in the presence of dust, $c_{\rm s}$ the adiabatic sound speed of gas, and $\xi=\Sigma_{\rm d,tot}/\Sigma_{\rm g}$ the total dust-to-gas ratio.
At the burst onset, the radial velocity  in the near-Keplerian disk  is small and is largely determined by the radial mass transport mechanisms operating in the disk (gravitational and viscous torques).  In the disk regions between 30 and 80~au occupied by strong spiral arms a local increase in the radial velocity is evident (see also the upper-left panel in Fig.~\ref{fig:11}). One hundred years after the onset of the burst, however, the radial velocity is predominantly positive and is notably increased throughout the disk (green line), reflecting the disk expansion caused by irradiation heating and gas pressure increase throughout the disk. After the burst, the disk quickly cools ( due to short thermal timescales) and starts shrinking again. The $Q$-parameter before the burst is everywhere greater than unity, indicating global stability of the disk to gravitational fragmentation. At the same time $Q<1.5$, meaning susceptibility to the development of gravitational instability in the form of spiral arms. At the end of the burst (500~yr), the disk  expands notably, and the $Q$-value becomes greater than 2 throughout the disk, resulting in disk transient stabilization (green and blue lines).  After the end of the burst, disk contraction ensues and causes the $Q$-parameter finally to drop below unity, triggering gravitational fragmentation (black line).

A similar expansion-contraction cycle causes disk fragmentation in the $t_{\rm bst}$=100~yr and $t_{\rm bst}$=200~yr models.
We repeated our simulations with a burst magnitude declining with time and found that disk fragmentation occurred in the $t_{\rm bst}=200$~yr model, but not in the $t_{\rm bst}=100$~yr and $t_{\rm bst}=500$~yr cases. Moreover, the burst-triggered disk fragmentation is not present at the other burst onset times at 245~kyr and 645~kyr.
We conclude that although the burst-triggered disk fragmentation is an intriguing effect, it is not expected to be widespread and may depend on the pre-burst conditions in the disk. 

\begin{figure}
\begin{centering}
\includegraphics[width=1\columnwidth]{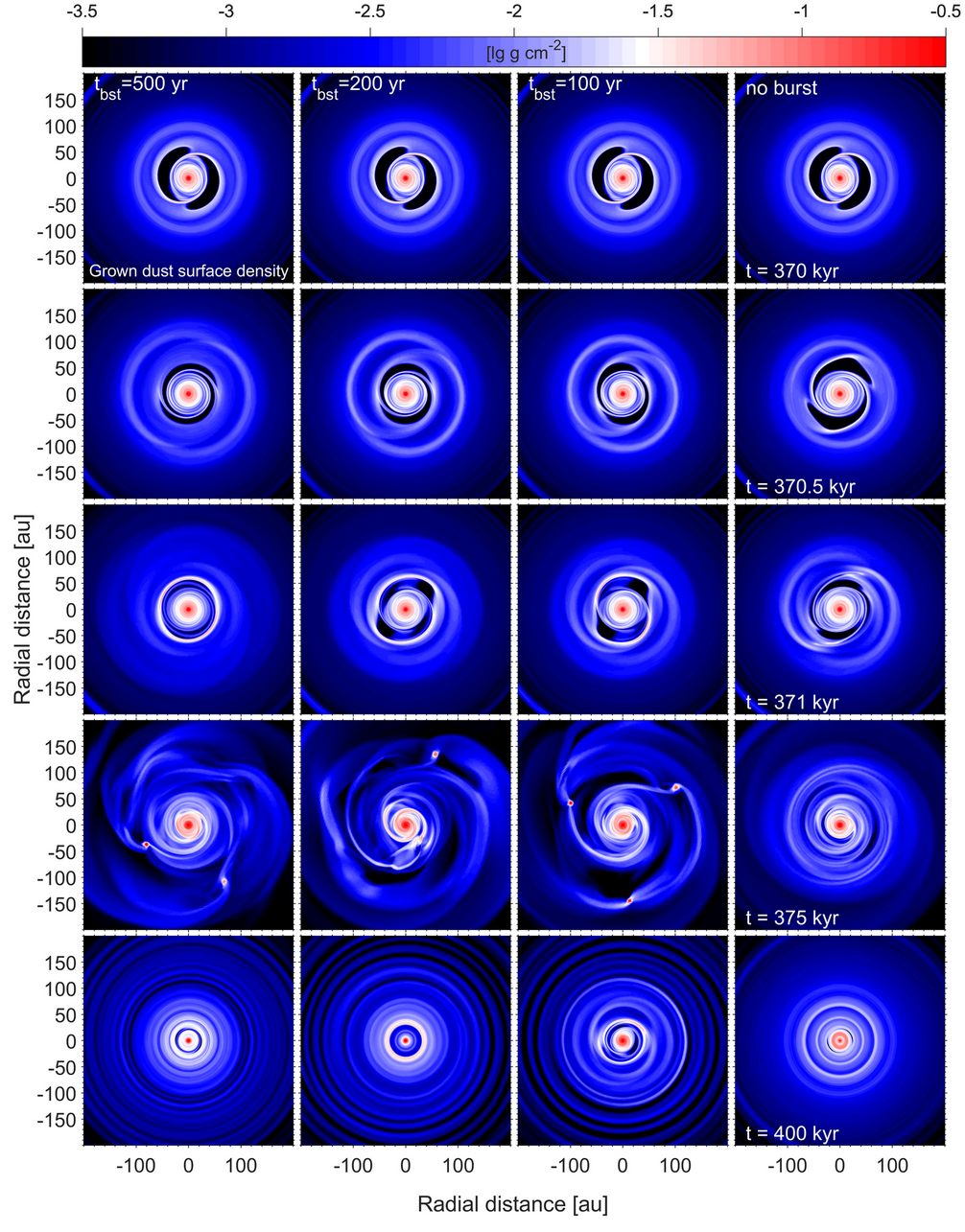}
\par\end{centering}
\caption{\label{fig:8} Similar to Figure~\ref{fig:5}, but for the grown dust.}
\end{figure}

Now we consider the long-term effect of the burst of various duration on the spatial distribution of grown dust. Figure~\ref{fig:8} presents the spatial distribution of  grown dust for the same burst characteristics and at the same time instances as in Figure~\ref{fig:5} showing the gas component. Each column corresponds to the model with a given $t_{\rm bst}$ as indicated in the top panels. The right-hand-side column presents the model without the burst. 
During most of the considered post-burst evolution the spatial distribution of grown dust is characterized by rings and deep gaps. The only exception is the time period of disk gravitational fragmentation when the disk is very dynamic and chaotic. 
The final ring morphology 30~kyr after the burst  is different in the burst and non-burst models as is evident from the bottom row of Figure~\ref{fig:8}. The burst models form multiple sharp rings at various radial distances. The wide-orbit rings likely result from dispersal of dust-rich clumps that form through disk fragmentation in the preceding strongly unstable phase. Indeed, gaseous clumps are known to collect dust as they drift through the gas disk \citep[e.g.,][]{2011ChaNayakshin,2019VorobyovElbakyan}. The model without burst also forms a ring-like structure, which is however different in appearance from those of the burst models. In particular, the number of the rings is smaller, and they all are located within 50~au. 
Whether of not our ring structures are potentially related to a variety of similar
structures observed in T~Tauri disks \citep{2015ALMA,2018Andrews,2018LongPinilla}
remains to be understood.
In a follow-up study, we plan to perform radiation transfer calculations on the obtained gas and dust structure to obtain dust intensity maps and compare them with the known cases. Finally, we note that the mass of grown dust in the rings is limited to several Earth masses, which may not be sufficient to form solid protoplanetary cores for dust-to-core conversion rates of up to 10\%.


\begin{figure}
\begin{centering}
\includegraphics[width=1\columnwidth]{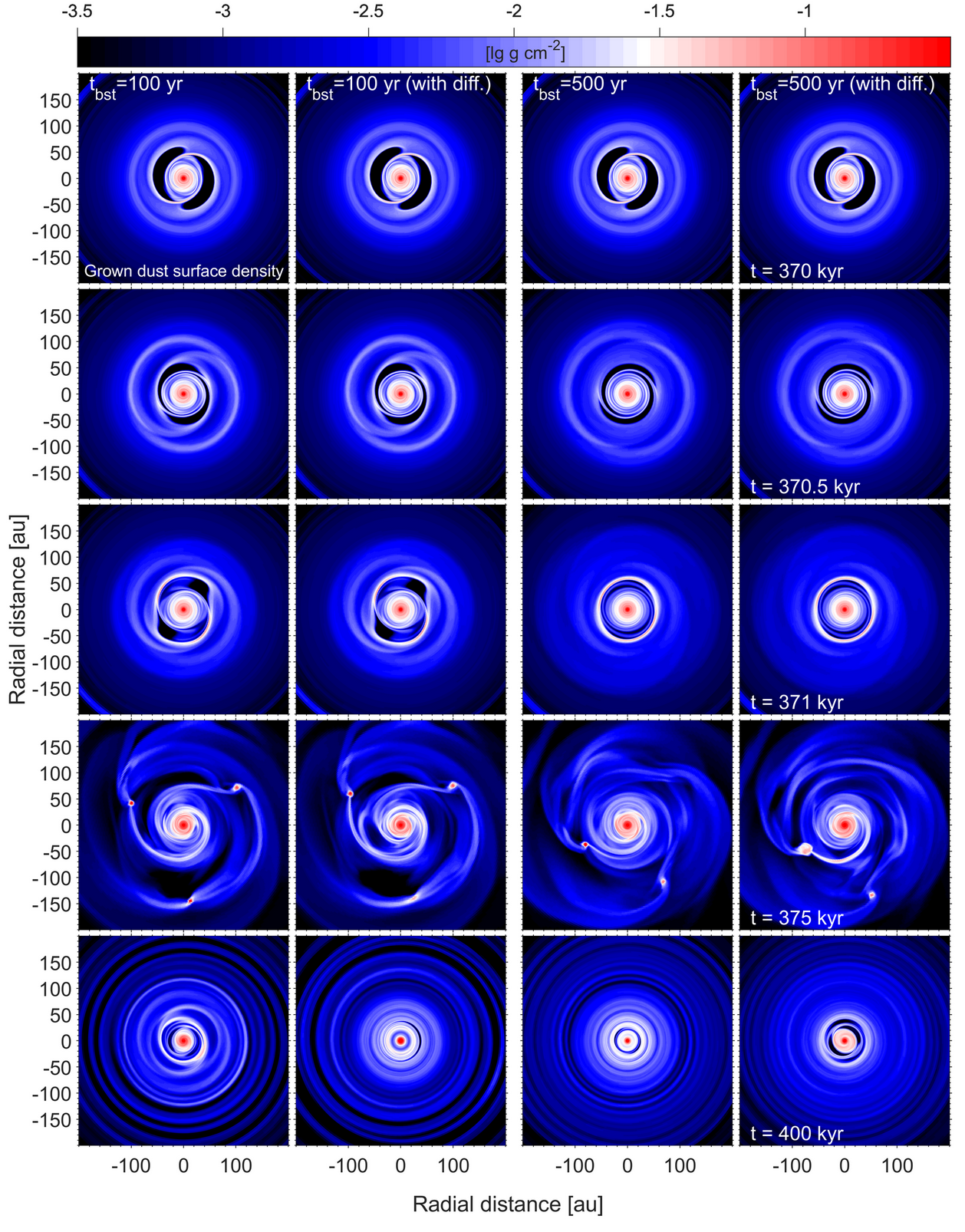}
\par\end{centering}
\caption{\label{fig:9} Effect of turbulent diffusion on the spatial distribution of grown dust. The two columns on the right present the comparison for the $t_{\rm bst}=500$~yr burst of constant amplitude, while the two columns on the left do that for the $t_{\rm bst}=100$~yr burst. The scale bar is in log g~cm$^{-2}$.}
\end{figure}

\section{The effect of dust diffusion}
The results for the dust distribution presented in previous sections did not take dust turbulent diffusion into account. It is well known that dust diffusion can smear out strong dust concentrations, thus potentially affecting the formation and longevity of the dust rings found in our numerical simulations. To test the effect, we re-run the models with constant luminosity bursts of 100~yr and 500~yr in duration and compared the resulting dust distributions with the original models without turbulent diffusion. To include the diffusion of dust, we modified the continuity equation~(\ref{contDlarge}) for grown dust as follows \citep{1988ClarkePringle}
\begin{equation}
  \label{contD_diff}
\frac{{\partial \Sigma_{\rm d,gr} }}{{\partial t}}  + \nabla_p  \cdot 
\left( \Sigma_{\rm d,gr} \bl{u}_p \right) = S(a_{\rm max}) + \nabla \cdot \left( D \Sigma_{\rm g} \nabla \left( {\Sigma_{\rm d,gr} \over \Sigma_{\rm g}} \right)  \right),  
\end{equation}
where $D$ is the turbulent diffusivity, which is related to the turbulent viscosity $\nu$ through the Schmidt number $D=\nu/Sc$. The value of $Sc$ is taken to be equal to unity, a typical and conservative choice when considering dust diffusion in protoplanetary disks \citep[e.g.][]{2012ZhuNelson,2015Dipierro}.  Clearly, the form of the diffusion term implies negligible diffusion when the distribution of dust exactly follows that of gas. Otherwise, dust density enhancements with respect to gas tend to be speared out. 

Figure~\ref{fig:9} presents the results of our test run. The two columns on the left show the grown dust surface density distributions without and with diffusion for $t_{\rm bst}=100$~yr, while the two columns on the right do that for $t_{\rm bst}=500$~yr. Dust diffusion has little effect on the distribution of grown dust on short timescales of the burst. This is not surprising since the viscous timescale $\tau_{\rm visc}=r^2/\nu$ is much shorter than the burst duration ($\tau_{\rm visc}\sim 10^4$~yr at $r=10$~au and $\tau_{\rm visc}\sim 10^5$~yr at $r=50$~au). On longer times on the order of tens of thousand years the effect of dust diffusion becomes notable as the bottom row in Figure~\ref{fig:9} demonstrates. The rings are less sharp and the arrangement of the rings is also somewhat affected by dust diffusion but the rings do not smear out completely.

\section{Summary}
\label{Summary}
In this paper, we studied numerically  the long-term evolution of a gas-dust disk and the response of a gravitationally unstable disk to accretion and luminosity bursts of various duration.
We employed numerical hydrodynamics simulations in the thin-disk limit using the FEOSAD code \citep{2018VorobyovAkimkin,2019VorobyovSkliarevskii}, which allowed us to study the evolution of both gas and dust disk subsystems including dust growth and dust-to-gas friction with back reaction. We initiate a burst that is consistent with triggering of the MRI via thermal ionization of alkaline metals in the innermost regions at temperatures exceeding 1300~K. Three bursts of various duration ($t_{\rm bst}$=100, 200, and 500~yr) were considered. In addition, two shapes of luminosity bursts were adopted: a constant-luminosity burst with a time-independent value of $100~L_\odot$ and a declining-luminosity burst with a peak value of $300~L_\odot$ gradually declining to $15~L_\odot$ at the end of the burst. Quiescent pre- and post-burst luminosities are on the order of $1~L_\odot$. The changes in the spatial distribution of both gas and grown dust were considered. The results of modeling were compared to the case without the burst. Our findings can be summarized as follows.


-- The long-term evolution of a protoplanetary disk can be characterized by recurrent episodes of disk gravitational destabilization, which result in the formation of a transient spiral pattern in the gas disk. This occurs if a dead zone is present in the innermost disk regions, causing accumulation of matter and triggering gravitational instability. The spatial distribution of grown dust (with the size that gradually declines from several centimeters in the inner 20~au to $\approx 1$~mm at 60~au) can be different from that of gas and can exhibit pronounced rings and gaps that gradually disappear as the spiral pattern in the gas disk diminishes. The spatial morphology of small sub-micron dust follows that of gas.


-- Bursts of all considered duration act to reduce the strength of gravitational instability in the gas disk by heating  the disk and causing it to expand. The effect is strongest for the longest burst with $t_{\rm bst}=500$~yr, for which the burst duration is comparable to the revolution period of the spiral and to the dynamical timescale at a distance where the spiral is most pronounced. In this case, the original two-armed spiral pattern in the gas disk completely diminishes by the end of the burst, leaving behind  only tracers  of  weak spiral  arms. The shortest-burst model with $t_{\rm bst}=100$~yr, on the other hand, can only weaken the spiral pattern in the gas disk due to short-lasting expansion, because the dynamical timescale and the revolution period of the spiral are much longer than the burst duration. 

-- The grown dust is found to reacts somewhat differently to the burst. The spiral-like initial distribution with deep cavities in the inter-armed regions (caused by dust drift  away from the inter-armed regions towards local pressure maxima in the arms) transforms into a ring-like distribution with deep gaps. This transformation is most pronounced for the burst of longest duration (500~yr). 

-- The long-term effect of the burst on the disk structure is versatile and may depends on the initial disk conditions at the onset of the burst. In some cases, the spiral structure recovers after the burst ends and even exceeds in strength the pre-burst level. Vigorous disk fragmentation sets in several thousand years after the burst, which was absent in the model without the burst. Several clumps with masses in the 1.0--7.0~$M_{\rm Jup}$ limits form in the outer disk regions, but are tidally destroyed soon after their formation. 

-- After the disk fragmentation phase ends, the spatial distribution of grown dust  is characterized by multiple rings located from tens to hundreds of astronomical units. The wide-orbit rings are likely formed as the result of dist-rich clump dispersal in gravitationally fragmenting disks.  Dust turbulent diffusion can affect the arrangement and sharpness of the rings but do not smear out them completely. 

The ring-like morphology of the spatial distribution of grown dust appears to be a typical outcome of long-duration luminosity bursts. But the rings are also found in models without bursts.  In particular, the ring morphology in models with bursts is different from that in models without bursts. Whether of not the obtained ring structures are related to a variety of rings and gaps observed in T~Tauri disks remains to be understood.

Our results may be useful in the context of V883~Ori, a FUor-like object that is thought to be in the outburst phase for more than a century with the unknown onset date.  The disk mass of V883~Ori is $\ga0.3~M_\odot$ \citep{2018Cieza}, suggesting that the disk may have been gravitationally unstable before the burst and may have possessed a developed spiral structure.  According to our simulations, a burst of 100–-200 yr in duration cannot fully destroy the spiral pattern in the gas disk. One needs a longer burst of several hundred years in duration. At the current stage of the V883~Ori outburst we may expect a notable weakening of its presumably preexisted spiral structure. Concurrently, the dust disk is expected to show a ring-like morphology with gaps. We lack high-resolution imaging of the V883~Ori's gas disk that would have allowed us to a make firm conclusions on the strength of gravitational instability in the gas disk, but the ring-like morphology detected in the millimeter emission of dust \citep{2016Cieza} is reminiscent of what we obtained in the longest outburst model with $t_{\rm bst}$=500~yr. Further observations and modeling with disk parameters closer matching those of V883~Ori are needed for interpreting its outburst.

\section*{Acknowledgements}
The authors are thankful to the anonymous referee for an insightful review that helped to improve the manuscript.
This work was supported by the Russian Fund for Fundamental Research, Russian-Taiwanese project 19-52-52011 and MoST project 108-2923-M-001-006-MY3.  The simulations were performed on the Vienna Scientific Cluster.  

\bibliographystyle{aa}
\bibliography{refs}

\end{document}